\documentclass[reprint,superscriptaddress,a4paper,aps,longbibliography]{revtex4-2}
\usepackage{graphicx}\graphicspath{{./}}
\usepackage{amsfonts}
\usepackage{revsymb}
\usepackage{amsmath}
\usepackage{amssymb}
\usepackage{bm}
\usepackage[usenames,dvipsnames]{xcolor}
\usepackage{oplotsymbl} 
\usepackage{fdsymbol} 
\DeclareMathAlphabet\mathbfcal{OMS}{cmsy}{b}{n}

\begin{document}
\title{Origin of the suppression of magnetic order in MnSi under hydrostatic pressure}
\author{P. Dalmas de R\'eotier}
\affiliation{Universit\'e Grenoble Alpes, CEA, Grenoble INP, IRIG-PHELIQS, F-38000 Grenoble, France}
\author{A. Yaouanc}
\affiliation{Universit\'e Grenoble Alpes, CEA, Grenoble INP, IRIG-PHELIQS, F-38000 Grenoble, France}
\author{D. Andreica}
\affiliation{Faculty of Physics, Babes-Bolyai University, 400084 Cluj-Napoca, Romania}
\author{R. Gupta}
\affiliation{Laboratory for Muon-Spin Spectroscopy, Paul Scherrer Institute,
  CH-5232 Villigen-PSI, Switzerland}
\affiliation{Department of Physics, Indian Institute of Technology Ropar, Rupnagar, Punjab 140001, India}
\author{R. Khasanov}
\affiliation{Laboratory for Muon-Spin Spectroscopy, Paul Scherrer Institute,
CH-5232 Villigen-PSI, Switzerland}
\author{G. Lapertot}
\affiliation{Universit\'e Grenoble Alpes, CEA, Grenoble INP, IRIG-PHELIQS, F-38000 Grenoble, France}

\date{\today}

\begin{abstract}

We experimentally study the evolution of the  magnetic moment $m$ and exchange interaction $J$ as a function of hydrostatic pressure in the zero-field helimagnetic phase of the strongly correlated electron system MnSi. The suppression of magnetic order at $\approx 1.5$~GPa is shown to arise from the  $J$ collapse  and not from a quantum fluctuations induced reduction of $m$. Our work provides benchmarks for first principles theories that are challenged by the presence of strong correlations and the possible role of Hund's coupling. In addition, our experimental data are consistent with a reorientation of the magnetic propagation wavevector recently evidenced above $\approx 1.2$~GPa. This result calls for a thorough investigation of the crystal structure in this pressure range.
\end{abstract}

\maketitle

\noindent{\sl Introduction} ---
Over the recent years the concept of quantum materials has been introduced to qualify a class of strongly correlated electron systems whose behavior escapes description within the framework of modern condensed-matter theories \cite{Fulde12}. They often exhibit emergent phenomena resulting from the interplay of charge, spin, orbit and lattice degrees of freedom \cite{NatPhys16}. Manganese silicide MnSi is one of the prominent systems in which strong electron correlations lead to exotic properties. This compound is also the first in which a magnetic skyrmion crystal was ever evidenced \cite{Muhlbauer09b}. This discovery triggered a tremendous interest owing to fundamental questions raised by the topological nature of the magnetic skyrmion texture and by potential applications for the storage of magnetic information and in spintronics.

With an ordered magnetic moment $m$ much smaller than the paramagnetic moment, and a relatively low magnetic ordering temperature $T_{\rm c}$, MnSi has been classified as a weak itinerant magnet. The self-consistent renormalization theory of the spin fluctuations \cite{Moriya85,Kakehashi13} successfully rationalizes the $T_{\rm c}$ value and the paramagnetic moment \cite{Lonzarich85}. However, the dual nature of MnSi with itinerant and localized electrons should be accounted for \cite{Ziebeck82,Fawcett89,Yaouanc20,Chen20,Choi19,Dalmas21}. 

{\sl Ab initio\/} theories and related calculations provide in principle an elegant framework to grasp the ground states and excitations of systems with intricate degrees of freedom, for which more conventional theories require the introduction of phenomenological parameters that are not related to each other. These methods have been successful for an insight into the properties of MnSi or similar systems; see, e.g.\ Refs.~\onlinecite{Grytsiuk19,Grytsiuk21,Borisov21,MendiveTapia21} for recent reports. However the strong longitudinal fluctuations of the magnetic moment of MnSi associated with the small value of $m$ are challenging for first principles theoretical studies \cite{Jeong04,Collyer08}. Local spin density approximation calculations conclude to a value consistent with $m$ only for a lattice parameter sizably smaller than the actual $a_{\rm latt} = 0.4558$~nm \cite{Lerch94,Jeong04}. More recent density functional theory (DFT) calculations still predict a lattice parameter smaller than observed \cite{Hortamani08}. The dual character of the electronic density and the need to account for the correlations between the two subsets in terms of Hund's metal \cite{Fang22} are additional challenges for the theory. A late work based on DFT combined with dynamical mean-field theory predicts a value for the Heisenberg exchange constant much larger than the experimental observation \cite{Borisov21,Dalmas24}. 

\begin{figure}
\includegraphics[width=\linewidth]{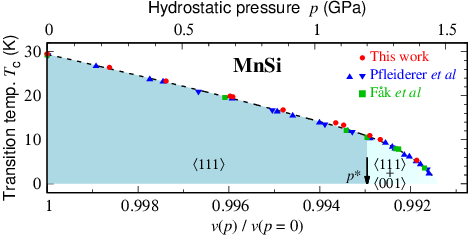}
\caption{Magnetic transition temperature as a function of the hydrostatic pressure or of the volume $v$ of the unit cell. The transition temperatures resulting from this work have been obtained from weak transverse-field $\mu$SR measurements; see Sect.~S2 of the Supplemental Material \cite{SM} for details. For comparison, results of resistivity ({\color{blue} $\blacktriangle$}), magnetic susceptibility ({\color{blue} $\blacktriangledown$}) \cite{Pfleiderer97}, and neutron scattering ({\color{ForestGreen} $\blacksquare$}) \cite{Fak05} experiments are also depicted. The shaded areas represent the different magnetic phases below and above $p^*\approx$~1.2~GPa, explicited in the main text and consistent with our observation. In both regions the sample is fully magnetic. The dashed line is a guide to the eye. The relation between pressure and volume is computed using the Murnaghan equation of state with experimentally measured values for the bulk modulus and its pressure derivative \cite{Brazhkin16}. The resulting volume variation is consistent with that of $a_{\rm latt}$ measured in neutron scattering experiments \cite{Fak05}. }
\label{Tc_p}
\end{figure}

Manganese silicide crystallizes in a so-called B20 cubic structure described by nonsymmorphic space group $P2_13$, with four manganese atoms in a unit cell \cite{Boren33}. The structure lacks a center of symmetry. In the presence of spin-orbit coupling, this authorizes the existence of the antisymmetric Dzyaloshinskii-Moriya (DM) interaction. Together with the dominant Heisenberg ferromagnetic exchange interaction, it is responsible for the helical magnetic structure observed in zero-field (ZF) below $T_{\rm c} \approx 30$~K \cite{Motoya76,Ishikawa76}. The phase transition has a weak-first order character \cite{Date77,*Stishov07}.
The evolution of $T_{\rm c}$ under hydrostatic pressure $p$ has been experimentally determined; see, e.g., Refs.~\onlinecite{Pfleiderer97} and \onlinecite{Fak05}. The phase diagram up to $p_{\rm c} \approx 1.5$~GPa is summarized in Fig.~\ref{Tc_p}, focusing on the magnetic order. From ambient pressure up to $p^* \approx 1.2$~GPa, the propagation wavevector ${\bf k}$ of the magnetic structure
is parallel to $\langle 111\rangle$ \cite{Ishikawa76,Fak05}. For $p^* \leq p \leq p_{\rm c}$,  ${\bf k}$ has recently been found to reorient along $\langle 001\rangle$ \cite{Bannenberg19}. In the whole pressure range up to $p_{\rm c}$, the system is fully magnetic \cite{Andreica10}. 

In this letter, through ZF muon spin rotation ($\mu$SR) measurements performed under hydrostatic pressure, we determine the dependence of the exchange interaction $J$ and of $m$ in MnSi with the volume $v$ of the unit cell and therefore with $a_{\rm latt}$. Our results provide novel constraints for the theories and prove that the suppression of the magnetic order at $p_{\rm c}$ is not due to quantum fluctuations or to the cancellation of $m$ but arises from the collapse of $J$. In addition the $\mu$SR spectra recorded for $p<p^*$ are fully consistent with the twisted and canted magnetic structure with ${\bf k}\parallel \langle 111\rangle$ established at low pressure \cite{Dalmas24}, while above $p^*$ they depart from the model, giving support to a recent neutron scattering result \cite{Bannenberg19}. 

The variation of $J$ is determined from the low temperature evolution of the spontaneous moment at different pressures using $\mu$SR measurements performed in the ordered phase of MnSi. This technique has proved to provide a very accurate measure of $m$ \cite{Dalmas16,Yaouanc20}, which is found to quadratically decay as a function of temperature, a behavior that we interpret in terms of the thermal excitation of helimagnons. The coefficient controlling the decrease of $m(T)$ allows us to infer $J$. The equation of state of MnSi is finally used to convert the pressure into the unit cell volume variation.

A seminal experimental investigation of a correlated electron compound under hydrostatic pressure was published about fifty years ago \cite{Huber75}. However, it is the discoveries around 2000 of the interplay of magnetism and superconductivity \cite{Mathur98,Saxena00} and induced Fermi liquid instabilities \cite{Lohneysen07}, both observed under few gigapascal hydrostatic pressure, that led to a sustained interest for these types of work. The vast majority of the investigations has been focused on the determination of phase diagrams from  transport, thermodynamic and magnetic bulk measurements. The number of reports discussing $J(p)$ is extremely limited. We are aware of neutron scattering studies of magnetic excitations concerning terbium metal \cite{Kawano94} and CsFeCl$_3$ \cite{Hayashida19}. Hydrostatic pressure has a very moderate effect in the first system while it reinforces the dominant exchange constant in the second.

\noindent{\sl Magnetic structure of MnSi} ---
The ZF magnetic structure of MnSi at ambient pressure has been established in several steps. Quasi-simultaneous neutron scattering and ZF nuclear magnetic resonance (NMR) measurements were interpreted in terms of a long-pitch helimagnetic structure \cite{Ishikawa76,Motoya76}. Lately, from a detailed analysis of a ZF-$\mu$SR spectrum, a double-helix structure was unveiled with one of the four manganese magnetic moments of the cubic unit cell drawing an helix along ${\bf k}$ as one moves from cell to cell, while the other three moments belong to a second helix twisted relative to the first one \cite{Dalmas16}. More recently, the data interpretation constrained to minimize the sum of  Heisenberg and DM energies concluded in the presence of a canting for the three aforementioned manganese magnetic moments in addition to the twist \cite{Dalmas24}.

The magnetic structure is fully consistent with the crystal structure symmetry, even when twisted and canted \cite{Dalmas16,Dalmas24}. To the best of our knowledge, no report exists of symmetry breaking of space group $P2_13$ under hydrostatic pressure. However $a_{\rm latt}$ extrapolated towards zero temperature, exhibits a singular evolution for $p^* \leq p \leq p_{\rm c}$: its magnetic and electronic contribution changes from positive for $p < p_{\rm c}$, corresponding to an expansion, to negative above $p_{\rm c}$, equivalent to a contraction \cite{Pfleiderer07}.

\noindent{\sl The model} --- We introduce the framework for the interpretation of our data. It starts with the spin Hamiltonian and the inference of the magnetic texture. The knowledge of the texture details is indeed required for the computation of the quantity measured in $\mu$SR experiments. In a second step, from the dispersion relation of the spin waves in the helimagnetic phase, we derive $m(T)$ and examine its dependence on $J$.

The spin Hamiltonian is written
\begin{eqnarray}
{\mathcal H}  & = &   - \frac {1}{2}
  \sum_{\langle{\bf i},{\bf i}^\prime,\gamma,{\gamma^\prime}\rangle}
  J\,{\bf S}_{{\bf i}, \gamma}\cdot {\bf S}_{{\bf i}^\prime, \gamma^\prime} \cr
 &  & + \frac {1}{2} \sum_{\langle{\bf i},{\bf i}^\prime, \gamma,{\gamma^\prime}\rangle}
   {\bf D}_{{\bf i},\gamma; {\bf i}^\prime, \gamma^\prime} \cdot
        \left ( {\bf S}_{{\bf i}, \gamma}\times {\bf S}_{{\bf i}^\prime,\gamma^\prime} \right ),
\label{Hamiltonian}
\end{eqnarray}
where ${\bf D}_{{\bf i},\gamma; {\bf i}^\prime, \gamma^\prime}$ denotes the Moriya vector associated with bond linking manganese sites $\gamma$ of unit cell ${\bf i}$ and $\gamma^\prime$ of cell ${\bf i}^\prime$. In Eq.~\ref{Hamiltonian} the sums are restricted to nearest neighbor spins ${\bf S}$. With four manganese atoms in a cubic unit cell, each of them having six manganese nearest neighbors, twenty-four ${\bf D}_{{\bf i},\gamma; {\bf i}^\prime, \gamma^\prime}$ vectors are to be considered. Accounting for the antisymmetry relation ${\bf D}_{{\bf i}^\prime, \gamma^\prime;{\bf i},\gamma} = -{\bf D}_{{\bf i},\gamma;{\bf i}^\prime,\gamma^\prime}$ and the symmetry elements of point group 23, the specification of the Moriya vector related to one of the bonds is sufficient to fix the other twenty-three. In practice we single out bond between sites ${\rm I}$ = $(x_{\rm Mn},x_{\rm Mn},x_{\rm Mn})$ and ${\rm II}$ = $(\bar x_{\rm Mn}+\frac{1}{2},\bar x_{\rm Mn}, x_{\rm Mn}+\frac{1}{2})$ in a unit cell of the cubic lattice and denote the related Moriya vector ${\bf D}_{{\bf i}, {\rm I};{\bf i},{\rm II}}$ as ${\bf D}$ \cite{SM}.

Allowing for a twisted and canted magnetic structure, the magnetic energy associated with $\mathcal H$ can be analytically minimized thanks to the incommensurate nature of the magnetic structure \cite{Chizhikov12}, with the minimum found for 
\begin{eqnarray} 
\sigma_1  \equiv \frac{D^x - D^y +2 D^z}{J} = \frac{3 k a_{\rm latt}}{2}.
\label{B1_1}
\end{eqnarray}
The twist and canting angles minimizing the energy can be expressed as a function of the single parameter 
\begin{eqnarray}
\sigma_2 \equiv \frac{D^x + D^y}{J}.
\label{B1_2}
\end{eqnarray}

The time scale of the $\mu$SR oscillations enables the determination of $m$. Its temperature dependence reflects the thermal excitation of helimagnons. From their dispersion relation \cite{Maleyev06,Belitz06} and using the same methodology as for the derivation of the $T^{3/2}$ Bloch's law for ferromagnets \cite{vanKranendonk58}, a quadratic decay of $m$ in temperature is found \cite{Yaouanc20,Dalmas21,SM},
\begin{eqnarray}
m(T) = m(T = 0) \left[1 - \left (\frac{T}{T_{\rm he}}\right )^2 \right],
\label{B1_3}
\end{eqnarray}
where parameter $T_{\rm he}$ is related to $J$ through the relation \cite{SM}
\begin{equation}
  J = \sqrt{\frac{\pi }{3 k\, a_{\rm latt}}} \frac{k_{\rm B} T_{\rm he}}{12\epsilon^{1/4}S}.
  \label{B1_4}
\end{equation}
Here $\epsilon$ is a constant, theoretically equal to 1/2 \cite{Maleyev06}. Experimentally, at zero pressure, $\epsilon =0.57 \, (5)$ \cite{Yaouanc20}, which is consistent with expectation. In the following, we will use Eqs.~\ref{B1_3} and \ref{B1_4} to determine $J(p)$ from accurate measurements of $m(T)$ at selected pressures.

\begin{figure}
\includegraphics[width=\linewidth]{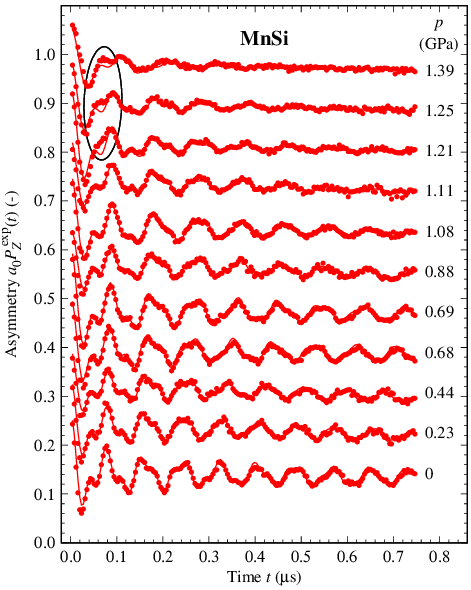}
\caption{Zero-field $\mu$SR spectra at various pressures. The temperature is 0.25~K except for the 0 and 0.69~GPa spectra which were measured at 5~K. The spectrum at $p$ = 0 has been recorded in a unpressurized CuBe cell for checking purpose. The spectra at $p>0$ are drawn with a vertical shift for clarity. The full lines depict the results of fits with the model described in the main text and Ref.~\onlinecite{SM}. Note the slight misfit around 0.06~$\mu$s for $p \ge 1.21$~GPa, which is discussed in the main text.}
\label{spectra_lowT}
\end{figure}

\noindent{\sl Experimental} ---
Two different cylindrical-shape single crystals and three types of piston-cylinder pressure cells have been used for the experiments. A 19~mm long and 7~mm diameter crystal \footnote{This sample was already used for the studies published in Refs.~\onlinecite{Andreica10,Amato14}.} has been  mounted in a CuBe pressure cell for measurements performed at 0.68 and 0.69~GPa and in a MP35N pressure cell for measurements at 0.23, 0.44, 0.88, and 1.11~GPa. Experiments with a second single crystal of 12~mm length and 6~mm diameter inserted in a double wall pressure cell with MP35N outer wall and CuBe inner wall \cite{Shermadini17} have been performed at 1.08, 1.21, 1.25 and 1.39~GPa. The transmitting medium has been Daphne 7373 oil for the three types of cells. The pressure has been determined from the detection of the superconducting transition of a small indium piece inserted in the pressure cell together with the MnSi single crystal, through ac magnetic susceptibility measurements. The uncertainty on the pressure is estimated to $\approx 0.02$~GPa. The $\mu$SR spectra have been recorded with the General Purpose Decay-channel (GPD) spectrometer \cite{Khasanov16,Khasanov22} located at the Swiss muon Source of the Paul Scherrer Institute, Villigen, Switzerland. At each pressure, the temperature of the magnetic transition was systematically determined through weak transverse-field measurements, see Fig.~\ref{Tc_p} and Ref.~\onlinecite{SM} for details. These measurements performed on the crystal under study provide an intrinsic measure of the pressure in addition to the superconducting transition of In. Figure~\ref{spectra_lowT} displays low temperature ZF-$\mu$SR spectra measured under different pressures. 
Further spectra recorded as a function of temperature at selected pressures are also available in Figs.~S2, S3, and S4 of Ref.~\onlinecite{SM}.

\noindent{\sl The fitting function} ---
The quantity measured in a ZF-$\mu$SR experiment is the so-called polarization function $P_Z(t)$ \cite{Yaouanc11,Blundell22,Amato24}. It reflects the evolution with time $t$ of the Cartesian $Z$ component of the positive muon spin ${\bf S}_\mu$ averaged over the $\approx 10^8$ particles implanted during an experiment. By convention $Z$ defines the direction of the polarization of the muon beam. The  ${\bf S}_\mu$ evolution is ruled by the Larmor equation $\frac{{\rm d}{\bf S}_\mu}{{\rm d}t}$ = $\gamma_\mu {\bf S}_\mu\times {\bf B}_{\rm loc}$, where $\gamma_\mu$ = 851.6~Mrad\,s$^{-1}$\,T$^{-1}$ is the muon gyromagnetic ratio and ${\bf B}_{\rm loc}$ the spontaneous local field at the interstitial site of muon implantation. This field is the sum of the dipole field arising from the localized manganese moments and the Fermi contact field associated with the density of polarized electrons at the muon site. The oscillations observed in Fig.~\ref{spectra_lowT} reflect the muon-spin precession in field ${\bf B}_{\rm loc}$ arising from the MnSi magnetic order.

The model function for a ZF spectrum is
\begin{eqnarray}
P_Z(t) & = & f P_Z^{\rm he}(t) + (1-f) P_Z^{\rm pc}(t),
\label{Pz}
\end{eqnarray}
where $P_Z^{\rm he}(t)$ and $P_Z^{\rm pc}(t)$ describe the contributions of muons implanted in the MnSi crystal and in the pressure cell, with respective weights $f$ and $1-f$. The material of the pressure cells being non-magnetic, $P_Z^{\rm pc}(t)$ is of the Kubo-Toyabe type with parameters depending on the type of pressure cell \cite{Khasanov16}. 

Thanks to a prior determination of the muon site and Fermi contact interaction magnitude \cite{Amato14,Dalmas18}, the function $P_Z^{\rm he}(t)$ valid for a twisted and canted helimagnetic phase can be computed. It depends on $m$, $a_{\rm latt}$, ${\bf k}$, and $\sigma_2$; see Ref.~\onlinecite{Dalmas24} for full details.

{\sl Results} ---
The fits of $P_Z(t)$ to the data are excellent as illustrated in Figs.~\ref{spectra_lowT}, S2, S3, and S4 \cite{SM}. Interestingly, a small but consistent misfit is noticed around 0.06~$\mu$s at 1.21~GPa and above, i.e.\ precisely in the pressure range where a reorientation of ${\bf k}$ from $\langle 111\rangle$ to $\langle 001\rangle$ has recently been observed \cite{Bannenberg19}. As shown in Fig.~S6 the spectra at 1.25~GPa can be accounted for assuming the coexistence of regions in the crystal with the two different ${\bf k}$ orientations.

\begin{figure}
\includegraphics[width=\linewidth]{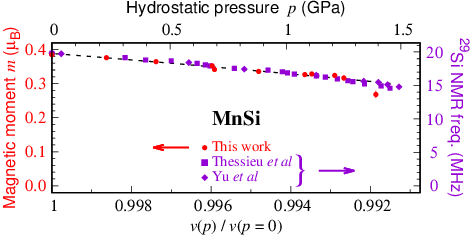}
\caption{Low temperature magnetic moment resulting from ZF-$\mu$SR measurements as a function of the volume of the unit cell. The spectra from which $m$ is determined were recorded at 0.25~K, except those at 0 and 0.69~GPa which were obtained at 5~K. For comparison the pressure dependence of the ZF $^{29}$Si nuclear magnetic resonance (NMR) frequency measured at 1.4~K (Thessieu {\em et al}\ \cite{Thessieu99}) and 1.8~K (Yu {\em et al}\/ \cite{Yu04}) is shown. We note that a resonance is detected up to 1.75 GPa in the latter reference. The dashed line is a guide to the eye.}
\label{m_p}
\end{figure}

\begin{figure}
\includegraphics[width=\linewidth]{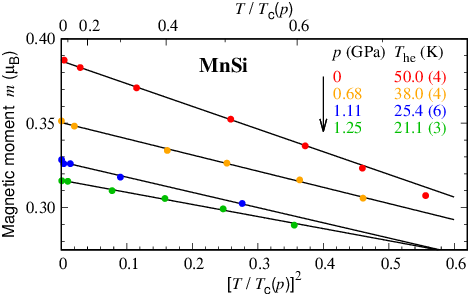}
\caption{Variation of the manganese magnetic moment as a function of the square of the normalized temperature at different pressures, as indicated in the graph. The full lines show the result of fits to Eq.~\ref{B1_3}. The refined $T_{\rm he}$ value is displayed next to each pressure. The 0~GPa dataset derives from fits of the model described in the main text to previously published $m(T)$ \cite{Yaouanc20}; see Sect.~S3B and Fig.~S5 \cite{SM}.}
\label{ZF_m_T2}
\end{figure}

The low temperature $m$ value at different pressures is plotted in Fig.~\ref{m_p}. The results match those of ZF-NMR.
The thermal variation of $m$ for four pressures is presented in Fig.~\ref{ZF_m_T2}. Here it must be emphasized that the coexistence of two magnetic regions in the crystal does not affect the $m$ values extracted from our fits; see Sect.~S4 for details \cite{SM}. At each of the pressures, $m(T)$ is well described by Eq.~\ref{B1_3}. 

\begin{figure}[tb]
\includegraphics[width=\linewidth]{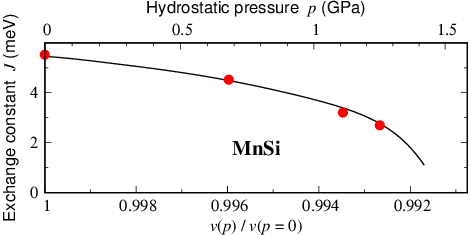}
\caption{Evolution of exchange constant $J$ with the volume of the unit cell.  Although most of the variation of $J$ follows from that of $T_{\rm he}$ (Eq.~\ref{B1_4}), the weak variations of $a_{\rm latt}$, $k$, and $S$ have been accounted for using Ref.~\onlinecite{Fak05} and Fig.~\ref{m_p}. The solid line is a fit with function $a T_{\rm c}(p)/m^2(p)$ which derives from the scaling law discussed in the main text.
The only free parameter is $a$.}
\label{B1_p}
\end{figure}

Accounting for  Eq.~\ref{B1_4}, we compute $J$ from $T_{\rm he}$, with results depicted in Fig.~\ref{B1_p}. The value $J(p=0) =  5.5 \, (1)$~meV, already reported in Ref.~\onlinecite{Yaouanc20}, compares well with $J \approx 6$~meV \footnote{The exchange $J$ is deduced from the spin wave dispersion $\hbar\omega({\bf q}) = D_{\rm sw}q^2$ valid for a ferromagnet, with $D_{\rm sw} = JSa_{\rm latt}^2$. The value for $J$ quoted in the main text is obtained for $D_{\rm sw}$ = 21\,(1)\,meV\,\AA$^2$ (Ref.~\onlinecite{Sato16}, see also Refs.~\onlinecite{Semadeni99,Boni11}) increased by $\approx 20\%$ to account for the fact that the measurements were performed at 27~K rather than at low temperature \cite{Ishikawa77}.} deduced from inelastic neutron scattering measurements of the spin wave dispersion \cite{Sato16,Boni11,Semadeni99} \footnote{We should note that contrasting values are found in the literature, e.g.\  $J$ = 11.4 or 21.8~meV \cite{Chen20,Jin23}. However they arise from studies focused at relatively large scattering vectors and energy transfers where the neutron intensities associated with the spin wave excitations and Stoner continuum interfere: this could explain the discrepancy with Refs.~\onlinecite{Sato16,Boni11,Semadeni99}.}. However, our prominent result is relative to the dependence $J(p)$. {\sl Figure~\ref{B1_p} strongly suggests that $J$ cancels at $p_{\rm c}$}, a result which can be interpreted as follows. From general spin-wave theory it is well established that $T_{\rm c} \propto J m^2$. This relation is quite general \cite{SM}: not only it is valid for instance in the mean-field approximation \cite{*[{See, e.g., }] [{}] Wang82} or in the random-phase approximation \cite{Bogolyubov59}, but it can also be derived from a simple scaling argument by equating the Heisenberg and thermal energies. Since $m(p)$ does not vanish while approaching $p_{\rm c}$ (Fig.~\ref{m_p}), in consistency with the first-order nature of the transition, the suppression of the magnetic order at $p_{\rm c}$ stems from the cancellation of the exchange interaction. 

\noindent{\sl Conclusions and perspectives} ---
From an accurate determination of $m(T)$ at different pressures, we have evidenced the dramatic $J(p)$ reduction when approaching $p_{\rm c}$, while $m(p)$ varies moderately. From the simple scaling relation between $T_{\rm c}$, $J$ and $m$, the disappearance of the long-range magnetic order at $p_{\rm c}$  is due to the $J(p_{\rm c})$ collapse rather than the $m(p_{\rm c})$ cancelling under the influence of quantum fluctuations \cite{Pfleiderer07,Povzner16}. The progressive reduction of $J$ supports the idea that the transition is only weakly first-order \cite{Petrova12}. It corroborates the absence of quantum fluctuations observed in MnSi (see, e.g., Ref.~\onlinecite{Pfleiderer07}) or in the solid solution Mn$_x$Fe$_{1-x}$Si \cite{Pappas21}. The suppression of the magnetic order under pressure in the sibling B20 compounds FeGe and MnGe could have the same origin as in MnSi, viz. the cancellation of $J$ and not that of $m$ \cite{Pedrazzini07,Deutsch14}. The Heisenberg interaction can naturally be attributed to the Ruderman--Kittel--Kasuya--Yosida mechanism \cite{Ruderman54,*Kasuya56,*Yosida57}. The itinerant electrons act as a medium through which an interaction is established between the electrons localized at manganese sites with a magnitude that depends on the distance between these sites and on the Fermi wavevector. In view of the modest evolution of $a_{\rm latt}$ in the considered pressure range, the Fermi surface is likely to be strongly affected by hydrostatic pressure. The concomitant instability in the itinerant electron system could be at the origin of the non-Fermi-liquid behavior in the paramagnetic state at low temperature \cite{Pfleiderer01,Doiron03,Wilde21}.

Our measurements show the twisted and canted ZF magnetic structure with ${\bf k} \parallel \langle 111\rangle$ to smoothly evolve up to $p^*$. There is no dramatic change in the ${\bf D}$ orientation. We observe a change above $p^*$, in consistency with the recently evidenced reorientation of ${\bf k}$ from $\langle 111\rangle$ to $\langle 001\rangle$ \cite{Bannenberg19}. The $\mu$SR spectra can be interpreted in terms of the coexistence of two regions, in line with the first order magnetic transition. The determination of the microscopic mechanism driving the ${\bf k}$ orientation in the crystal structure is urgently needed. The emergence of a region with a different ${\bf k}$ vector for $p^* \leq p \leq p_{\rm c}$ could be the consequence of a subtle crystallography change in this pressure range.

Our results provide benchmarks for the first-principle theories that are defied by some parameters measured on the B20 compounds, e.g., $J$ at ambient pressure is overestimated in MnSi \cite{Borisov21} or the computed helix pitch of FeGe is twice as large as observed \cite{Grytsiuk19}. The importance of Hund's coupling has been unveiled in recent {\sl ab-initio} works performed for related systems \cite{Gendron22,Cao24}. It would be of interest to consider this coupling for MnSi.

\begin{acknowledgments}
We acknowledge fruitful discussions with B.\ F{\aa}k about the physics of MnSi. We are grateful to K.\ Mony for assistance in the sample preparation. The $\mu$SR measurements were performed at the GPD spectrometer of the Swiss Muon Source (Paul Scherrer Institut, Villigen, Switzerland).
\end{acknowledgments}

\bibliography{reference,MnSi_J_p}

\begin{thebibliography}{85}%
\makeatletter
\providecommand \@ifxundefined [1]{%
 \@ifx{#1\undefined}
}%
\providecommand \@ifnum [1]{%
 \ifnum #1\expandafter \@firstoftwo
 \else \expandafter \@secondoftwo
 \fi
}%
\providecommand \@ifx [1]{%
 \ifx #1\expandafter \@firstoftwo
 \else \expandafter \@secondoftwo
 \fi
}%
\providecommand \natexlab [1]{#1}%
\providecommand \enquote  [1]{``#1''}%
\providecommand \bibnamefont  [1]{#1}%
\providecommand \bibfnamefont [1]{#1}%
\providecommand \citenamefont [1]{#1}%
\providecommand \href@noop [0]{\@secondoftwo}%
\providecommand \href [0]{\begingroup \@sanitize@url \@href}%
\providecommand \@href[1]{\@@startlink{#1}\@@href}%
\providecommand \@@href[1]{\endgroup#1\@@endlink}%
\providecommand \@sanitize@url [0]{\catcode `\\12\catcode `\$12\catcode
  `\&12\catcode `\#12\catcode `\^12\catcode `\_12\catcode `\%12\relax}%
\providecommand \@@startlink[1]{}%
\providecommand \@@endlink[0]{}%
\providecommand \url  [0]{\begingroup\@sanitize@url \@url }%
\providecommand \@url [1]{\endgroup\@href {#1}{\urlprefix }}%
\providecommand \urlprefix  [0]{URL }%
\providecommand \Eprint [0]{\href }%
\providecommand \doibase [0]{https://doi.org/}%
\providecommand \selectlanguage [0]{\@gobble}%
\providecommand \bibinfo  [0]{\@secondoftwo}%
\providecommand \bibfield  [0]{\@secondoftwo}%
\providecommand \translation [1]{[#1]}%
\providecommand \BibitemOpen [0]{}%
\providecommand \bibitemStop [0]{}%
\providecommand \bibitemNoStop [0]{.\EOS\space}%
\providecommand \EOS [0]{\spacefactor3000\relax}%
\providecommand \BibitemShut  [1]{\csname bibitem#1\endcsname}%
\let\auto@bib@innerbib\@empty
\bibitem [{\citenamefont {Fulde}(2012)}]{Fulde12}%
  \BibitemOpen
  \bibfield  {author} {\bibinfo {author} {\bibfnamefont {P.}~\bibnamefont
  {Fulde}},\ }\href@noop {} {\emph {\bibinfo {title} {Correlated Electrons in
  Quantum Matter}}}\ (\bibinfo  {publisher} {World Scientific},\ \bibinfo
  {address} {Singapore},\ \bibinfo {year} {2012})\BibitemShut {NoStop}%
\bibitem [{Nat(2016)}]{NatPhys16}%
  \BibitemOpen
  \bibfield  {title} {\bibinfo {title} {The rise of quantum materials},\ }\href
  {https://doi.org/10.1038/nphys3668} {\bibfield  {journal} {\bibinfo
  {journal} {Nat. Phys.}\ }\textbf {\bibinfo {volume} {12}},\ \bibinfo {pages}
  {105} (\bibinfo {year} {2016})}\BibitemShut {NoStop}%
\bibitem [{\citenamefont {M\"uhlbauer}\ \emph {et~al.}(2009)\citenamefont
  {M\"uhlbauer}, \citenamefont {Binz}, \citenamefont {Jonietz}, \citenamefont
  {Pfleiderer}, \citenamefont {Rosch}, \citenamefont {Neubauer}, \citenamefont
  {Georgii},\ and\ \citenamefont {B\"oni}}]{Muhlbauer09b}%
  \BibitemOpen
  \bibfield  {author} {\bibinfo {author} {\bibfnamefont {S.}~\bibnamefont
  {M\"uhlbauer}}, \bibinfo {author} {\bibfnamefont {B.}~\bibnamefont {Binz}},
  \bibinfo {author} {\bibfnamefont {F.}~\bibnamefont {Jonietz}}, \bibinfo
  {author} {\bibfnamefont {C.}~\bibnamefont {Pfleiderer}}, \bibinfo {author}
  {\bibfnamefont {A.}~\bibnamefont {Rosch}}, \bibinfo {author} {\bibfnamefont
  {A.}~\bibnamefont {Neubauer}}, \bibinfo {author} {\bibfnamefont
  {R.}~\bibnamefont {Georgii}},\ and\ \bibinfo {author} {\bibfnamefont
  {P.}~\bibnamefont {B\"oni}},\ }\bibfield  {title} {\bibinfo {title} {Skyrmion
  lattice in a chiral magnet},\ }\href
  {https://doi.org/10.1126/science.1166767} {\bibfield  {journal} {\bibinfo
  {journal} {Science}\ }\textbf {\bibinfo {volume} {323}},\ \bibinfo {pages}
  {915} (\bibinfo {year} {2009})}\BibitemShut {NoStop}%
\bibitem [{\citenamefont {Moriya}(1985)}]{Moriya85}%
  \BibitemOpen
  \bibfield  {author} {\bibinfo {author} {\bibfnamefont {T.}~\bibnamefont
  {Moriya}},\ }\href@noop {} {\emph {\bibinfo {title} {Spin fluctuations in
  itinerant electron magnetism}}}\ (\bibinfo  {publisher} {Springer},\ \bibinfo
  {address} {Berlin},\ \bibinfo {year} {1985})\BibitemShut {NoStop}%
\bibitem [{\citenamefont {Kakehashi}(2013)}]{Kakehashi13}%
  \BibitemOpen
  \bibfield  {author} {\bibinfo {author} {\bibfnamefont {Y.}~\bibnamefont
  {Kakehashi}},\ }\href@noop {} {\emph {\bibinfo {title} {Modern theory of
  magnetism in metals and alloys}}}\ (\bibinfo  {publisher} {Springer},\
  \bibinfo {address} {Berlin},\ \bibinfo {year} {2013})\BibitemShut {NoStop}%
\bibitem [{\citenamefont {Lonzarich}\ and\ \citenamefont
  {Taillefer}(1985)}]{Lonzarich85}%
  \BibitemOpen
  \bibfield  {author} {\bibinfo {author} {\bibfnamefont {G.~G.}\ \bibnamefont
  {Lonzarich}}\ and\ \bibinfo {author} {\bibfnamefont {L.}~\bibnamefont
  {Taillefer}},\ }\bibfield  {title} {\bibinfo {title} {Effect of spin
  fluctuations on the magnetic equation of state of ferromagnetic or nearly
  ferromagnetic metals},\ }\href {https://doi.org/10.1088/0022-3719/18/22/017}
  {\bibfield  {journal} {\bibinfo  {journal} {J. Phys. C: Solid State Phys.}\
  }\textbf {\bibinfo {volume} {18}},\ \bibinfo {pages} {4339} (\bibinfo {year}
  {1985})}\BibitemShut {NoStop}%
\bibitem [{\citenamefont {Ziebeck}\ \emph {et~al.}(1982)\citenamefont
  {Ziebeck}, \citenamefont {Capellmann}, \citenamefont {Brown},\ and\
  \citenamefont {Booth}}]{Ziebeck82}%
  \BibitemOpen
  \bibfield  {author} {\bibinfo {author} {\bibfnamefont {K.~R.~A.}\
  \bibnamefont {Ziebeck}}, \bibinfo {author} {\bibfnamefont {H.}~\bibnamefont
  {Capellmann}}, \bibinfo {author} {\bibfnamefont {P.~J.}\ \bibnamefont
  {Brown}},\ and\ \bibinfo {author} {\bibfnamefont {J.~G.}\ \bibnamefont
  {Booth}},\ }\bibfield  {title} {\bibinfo {title} {Spin fluctuations in both
  the ordered and paramagnetic phases of \uppercase{M}n\uppercase{S}i!
  \uppercase{M}n\uppercase{S}i a heavy \uppercase{F}ermi liquid?},\ }\href
  {https://doi.org/10.1007/BF01420586} {\bibfield  {journal} {\bibinfo
  {journal} {Z. Phys. B}\ }\textbf {\bibinfo {volume} {48}},\ \bibinfo {pages}
  {241} (\bibinfo {year} {1982})}\BibitemShut {NoStop}%
\bibitem [{\citenamefont {Fawcett}(1990)}]{Fawcett89}%
  \BibitemOpen
  \bibfield  {author} {\bibinfo {author} {\bibfnamefont {E.}~\bibnamefont
  {Fawcett}},\ }\bibfield  {title} {\bibinfo {title} {Magnetic {G}r\"uneisen
  parameters of critical fluctuations in spin-density wave metals {C}r,
  {M}n{S}i and {M}n$_3${S}i},\ }\href
  {https://doi.org/https://doi.org/10.1016/0921-4526(89)90109-9} {\bibfield
  {journal} {\bibinfo  {journal} {Physica B}\ }\textbf {\bibinfo {volume}
  {161}},\ \bibinfo {pages} {83} (\bibinfo {year} {1990})}\BibitemShut
  {NoStop}%
\bibitem [{\citenamefont {Yaouanc}\ \emph {et~al.}(2020)\citenamefont
  {Yaouanc}, \citenamefont {Dalmas~de R\'eotier}, \citenamefont {Roessli},
  \citenamefont {Maisuradze}, \citenamefont {Amato}, \citenamefont {Andreica},\
  and\ \citenamefont {Lapertot}}]{Yaouanc20}%
  \BibitemOpen
  \bibfield  {author} {\bibinfo {author} {\bibfnamefont {A.}~\bibnamefont
  {Yaouanc}}, \bibinfo {author} {\bibfnamefont {P.}~\bibnamefont {Dalmas~de
  R\'eotier}}, \bibinfo {author} {\bibfnamefont {B.}~\bibnamefont {Roessli}},
  \bibinfo {author} {\bibfnamefont {A.}~\bibnamefont {Maisuradze}}, \bibinfo
  {author} {\bibfnamefont {A.}~\bibnamefont {Amato}}, \bibinfo {author}
  {\bibfnamefont {D.}~\bibnamefont {Andreica}},\ and\ \bibinfo {author}
  {\bibfnamefont {G.}~\bibnamefont {Lapertot}},\ }\bibfield  {title} {\bibinfo
  {title} {Dual nature of magnetism in \uppercase{M}n\uppercase{S}i},\ }\href
  {https://doi.org/10.1103/PhysRevResearch.2.013029} {\bibfield  {journal}
  {\bibinfo  {journal} {Phys. Rev. Res.}\ }\textbf {\bibinfo {volume} {2}},\
  \bibinfo {pages} {013029} (\bibinfo {year} {2020})}\BibitemShut {NoStop}%
\bibitem [{\citenamefont {Chen}\ \emph {et~al.}(2020)\citenamefont {Chen},
  \citenamefont {Krivenko}, \citenamefont {Stone}, \citenamefont {Kolesnikov},
  \citenamefont {Wolf}, \citenamefont {Reznik}, \citenamefont {Bedell},
  \citenamefont {Lechermann},\ and\ \citenamefont {Wilson}}]{Chen20}%
  \BibitemOpen
  \bibfield  {author} {\bibinfo {author} {\bibfnamefont {X.}~\bibnamefont
  {Chen}}, \bibinfo {author} {\bibfnamefont {I.}~\bibnamefont {Krivenko}},
  \bibinfo {author} {\bibfnamefont {M.~B.}\ \bibnamefont {Stone}}, \bibinfo
  {author} {\bibfnamefont {A.~I.}\ \bibnamefont {Kolesnikov}}, \bibinfo
  {author} {\bibfnamefont {T.}~\bibnamefont {Wolf}}, \bibinfo {author}
  {\bibfnamefont {D.}~\bibnamefont {Reznik}}, \bibinfo {author} {\bibfnamefont
  {K.~S.}\ \bibnamefont {Bedell}}, \bibinfo {author} {\bibfnamefont
  {F.}~\bibnamefont {Lechermann}},\ and\ \bibinfo {author} {\bibfnamefont
  {S.~D.}\ \bibnamefont {Wilson}},\ }\bibfield  {title} {\bibinfo {title}
  {Unconventional \uppercase{H}und metal in a weak itinerant ferromagnet},\
  }\href {https://doi.org/10.1038/s41467-020-16868-4} {\bibfield  {journal}
  {\bibinfo  {journal} {Nat. Commun.}\ }\textbf {\bibinfo {volume} {11}},\
  \bibinfo {pages} {3076} (\bibinfo {year} {2020})}\BibitemShut {NoStop}%
\bibitem [{\citenamefont {Choi}\ \emph {et~al.}(2019)\citenamefont {Choi},
  \citenamefont {Tai},\ and\ \citenamefont {Zhu}}]{Choi19}%
  \BibitemOpen
  \bibfield  {author} {\bibinfo {author} {\bibfnamefont {H.}~\bibnamefont
  {Choi}}, \bibinfo {author} {\bibfnamefont {Y.-Y.}\ \bibnamefont {Tai}},\ and\
  \bibinfo {author} {\bibfnamefont {J.-X.}\ \bibnamefont {Zhu}},\ }\bibfield
  {title} {\bibinfo {title} {Spin-fermion model for skyrmions in
  \uppercase{M}n\uppercase{G}e derived from strong correlations},\ }\href
  {https://doi.org/10.1103/PhysRevB.99.134437} {\bibfield  {journal} {\bibinfo
  {journal} {Phys. Rev. B}\ }\textbf {\bibinfo {volume} {99}},\ \bibinfo
  {pages} {134437} (\bibinfo {year} {2019})}\BibitemShut {NoStop}%
\bibitem [{\citenamefont {{Dalmas de R\'eotier}}\ and\ \citenamefont
  {Yaouanc}(2021)}]{Dalmas21}%
  \BibitemOpen
  \bibfield  {author} {\bibinfo {author} {\bibfnamefont {P.}~\bibnamefont
  {{Dalmas de R\'eotier}}}\ and\ \bibinfo {author} {\bibfnamefont
  {A.}~\bibnamefont {Yaouanc}},\ }\bibfield  {title} {\bibinfo {title}
  {Zero-field $^{29}$\uppercase{S}i nuclear magnetic resonance signature of
  helimagnons in \uppercase{M}n\uppercase{S}i},\ }\href
  {https://doi.org/10.1016/j.jmmm.2021.168086} {\bibfield  {journal} {\bibinfo
  {journal} {J. Magn. Magn. Mater.}\ }\textbf {\bibinfo {volume} {537}},\
  \bibinfo {pages} {168086} (\bibinfo {year} {2021})}\BibitemShut {NoStop}%
\bibitem [{\citenamefont {Grytsiuk}\ \emph {et~al.}(2019)\citenamefont
  {Grytsiuk}, \citenamefont {Hoffmann}, \citenamefont {Hanke}, \citenamefont
  {Mavropoulos}, \citenamefont {Mokrousov}, \citenamefont {Bihlmayer},\ and\
  \citenamefont {Bl\"ugel}}]{Grytsiuk19}%
  \BibitemOpen
  \bibfield  {author} {\bibinfo {author} {\bibfnamefont {S.}~\bibnamefont
  {Grytsiuk}}, \bibinfo {author} {\bibfnamefont {M.}~\bibnamefont {Hoffmann}},
  \bibinfo {author} {\bibfnamefont {J.-P.}\ \bibnamefont {Hanke}}, \bibinfo
  {author} {\bibfnamefont {P.}~\bibnamefont {Mavropoulos}}, \bibinfo {author}
  {\bibfnamefont {Y.}~\bibnamefont {Mokrousov}}, \bibinfo {author}
  {\bibfnamefont {G.}~\bibnamefont {Bihlmayer}},\ and\ \bibinfo {author}
  {\bibfnamefont {S.}~\bibnamefont {Bl\"ugel}},\ }\bibfield  {title} {\bibinfo
  {title} {Ab initio analysis of magnetic properties of the prototype
  \uppercase{B}20 chiral magnet \uppercase{F}e\uppercase{G}e},\ }\href
  {https://doi.org/10.1103/PhysRevB.100.214406} {\bibfield  {journal} {\bibinfo
   {journal} {Phys. Rev. B}\ }\textbf {\bibinfo {volume} {100}},\ \bibinfo
  {pages} {214406} (\bibinfo {year} {2019})}\BibitemShut {NoStop}%
\bibitem [{\citenamefont {Grytsiuk}\ and\ \citenamefont
  {Bl\"ugel}(2021)}]{Grytsiuk21}%
  \BibitemOpen
  \bibfield  {author} {\bibinfo {author} {\bibfnamefont {S.}~\bibnamefont
  {Grytsiuk}}\ and\ \bibinfo {author} {\bibfnamefont {S.}~\bibnamefont
  {Bl\"ugel}},\ }\bibfield  {title} {\bibinfo {title} {Micromagnetic
  description of twisted spin spirals in the \uppercase{B}20 chiral magnet
  \uppercase{F}e\uppercase{G}e from first principles},\ }\href
  {https://doi.org/10.1103/PhysRevB.104.064420} {\bibfield  {journal} {\bibinfo
   {journal} {Phys. Rev. B}\ }\textbf {\bibinfo {volume} {104}},\ \bibinfo
  {pages} {064420} (\bibinfo {year} {2021})}\BibitemShut {NoStop}%
\bibitem [{\citenamefont {Borisov}\ \emph {et~al.}(2021)\citenamefont
  {Borisov}, \citenamefont {Kvashnin}, \citenamefont {Ntallis}, \citenamefont
  {Thonig}, \citenamefont {Thunstr\"om}, \citenamefont {Pereiro}, \citenamefont
  {Bergman}, \citenamefont {Sj\"oqvist}, \citenamefont {Delin}, \citenamefont
  {Nordstr\"om},\ and\ \citenamefont {Eriksson}}]{Borisov21}%
  \BibitemOpen
  \bibfield  {author} {\bibinfo {author} {\bibfnamefont {V.}~\bibnamefont
  {Borisov}}, \bibinfo {author} {\bibfnamefont {Y.~O.}\ \bibnamefont
  {Kvashnin}}, \bibinfo {author} {\bibfnamefont {N.}~\bibnamefont {Ntallis}},
  \bibinfo {author} {\bibfnamefont {D.}~\bibnamefont {Thonig}}, \bibinfo
  {author} {\bibfnamefont {P.}~\bibnamefont {Thunstr\"om}}, \bibinfo {author}
  {\bibfnamefont {M.}~\bibnamefont {Pereiro}}, \bibinfo {author} {\bibfnamefont
  {A.}~\bibnamefont {Bergman}}, \bibinfo {author} {\bibfnamefont
  {E.}~\bibnamefont {Sj\"oqvist}}, \bibinfo {author} {\bibfnamefont
  {A.}~\bibnamefont {Delin}}, \bibinfo {author} {\bibfnamefont
  {L.}~\bibnamefont {Nordstr\"om}},\ and\ \bibinfo {author} {\bibfnamefont
  {O.}~\bibnamefont {Eriksson}},\ }\bibfield  {title} {\bibinfo {title}
  {Heisenberg and anisotropic exchange interactions in magnetic materials with
  correlated electronic structure and significant spin-orbit coupling},\ }\href
  {https://doi.org/10.1103/PhysRevB.103.174422} {\bibfield  {journal} {\bibinfo
   {journal} {Phys. Rev. B}\ }\textbf {\bibinfo {volume} {103}},\ \bibinfo
  {pages} {174422} (\bibinfo {year} {2021})}\BibitemShut {NoStop}%
\bibitem [{\citenamefont {Mendive-Tapia}\ \emph {et~al.}(2021)\citenamefont
  {Mendive-Tapia}, \citenamefont {dos Santos~Dias}, \citenamefont {Grytsiuk},
  \citenamefont {Staunton}, \citenamefont {Bl\"ugel},\ and\ \citenamefont
  {Lounis}}]{MendiveTapia21}%
  \BibitemOpen
  \bibfield  {author} {\bibinfo {author} {\bibfnamefont {E.}~\bibnamefont
  {Mendive-Tapia}}, \bibinfo {author} {\bibfnamefont {M.}~\bibnamefont {dos
  Santos~Dias}}, \bibinfo {author} {\bibfnamefont {S.}~\bibnamefont
  {Grytsiuk}}, \bibinfo {author} {\bibfnamefont {J.~B.}\ \bibnamefont
  {Staunton}}, \bibinfo {author} {\bibfnamefont {S.}~\bibnamefont {Bl\"ugel}},\
  and\ \bibinfo {author} {\bibfnamefont {S.}~\bibnamefont {Lounis}},\
  }\bibfield  {title} {\bibinfo {title} {Short period magnetization texture of
  {B}20-{M}n{G}e explained by thermally fluctuating local moments},\ }\href
  {https://doi.org/10.1103/PhysRevB.103.024410} {\bibfield  {journal} {\bibinfo
   {journal} {Phys. Rev. B}\ }\textbf {\bibinfo {volume} {103}},\ \bibinfo
  {pages} {024410} (\bibinfo {year} {2021})}\BibitemShut {NoStop}%
\bibitem [{\citenamefont {Jeong}\ and\ \citenamefont
  {Pickett}(2004)}]{Jeong04}%
  \BibitemOpen
  \bibfield  {author} {\bibinfo {author} {\bibfnamefont {T.}~\bibnamefont
  {Jeong}}\ and\ \bibinfo {author} {\bibfnamefont {W.~E.}\ \bibnamefont
  {Pickett}},\ }\bibfield  {title} {\bibinfo {title} {Implications of the {B}20
  crystal structure for the magnetoelectronic structure of {M}n{S}i},\ }\href
  {https://doi.org/10.1103/PhysRevB.70.075114} {\bibfield  {journal} {\bibinfo
  {journal} {Phys. Rev. B}\ }\textbf {\bibinfo {volume} {70}},\ \bibinfo
  {pages} {075114} (\bibinfo {year} {2004})}\BibitemShut {NoStop}%
\bibitem [{\citenamefont {Collyer}\ and\ \citenamefont
  {Browne}(2008)}]{Collyer08}%
  \BibitemOpen
  \bibfield  {author} {\bibinfo {author} {\bibfnamefont {R.}~\bibnamefont
  {Collyer}}\ and\ \bibinfo {author} {\bibfnamefont {D.}~\bibnamefont
  {Browne}},\ }\bibfield  {title} {\bibinfo {title} {Correlations and the
  magnetic moment of {M}n{S}i},\ }\href
  {https://doi.org/https://doi.org/10.1016/j.physb.2007.10.301} {\bibfield
  {journal} {\bibinfo  {journal} {Physica B}\ }\textbf {\bibinfo {volume}
  {403}},\ \bibinfo {pages} {1420} (\bibinfo {year} {2008})}\BibitemShut
  {NoStop}%
\bibitem [{\citenamefont {Lerch}\ and\ \citenamefont
  {Jarlborg}(1994)}]{Lerch94}%
  \BibitemOpen
  \bibfield  {author} {\bibinfo {author} {\bibfnamefont {P.}~\bibnamefont
  {Lerch}}\ and\ \bibinfo {author} {\bibfnamefont {T.}~\bibnamefont
  {Jarlborg}},\ }\bibfield  {title} {\bibinfo {title} {Magnetic and structural
  properties of {M}n{S}i},\ }\href
  {https://doi.org/https://doi.org/10.1016/0304-8853(94)90275-5} {\bibfield
  {journal} {\bibinfo  {journal} {J. Mag. Magn. Mat.}\ }\textbf {\bibinfo
  {volume} {131}},\ \bibinfo {pages} {321} (\bibinfo {year}
  {1994})}\BibitemShut {NoStop}%
\bibitem [{\citenamefont {Hortamani}\ \emph {et~al.}(2008)\citenamefont
  {Hortamani}, \citenamefont {Sandratskii}, \citenamefont {Kratzer},
  \citenamefont {Mertig},\ and\ \citenamefont {Scheffler}}]{Hortamani08}%
  \BibitemOpen
  \bibfield  {author} {\bibinfo {author} {\bibfnamefont {M.}~\bibnamefont
  {Hortamani}}, \bibinfo {author} {\bibfnamefont {L.}~\bibnamefont
  {Sandratskii}}, \bibinfo {author} {\bibfnamefont {P.}~\bibnamefont
  {Kratzer}}, \bibinfo {author} {\bibfnamefont {I.}~\bibnamefont {Mertig}},\
  and\ \bibinfo {author} {\bibfnamefont {M.}~\bibnamefont {Scheffler}},\
  }\bibfield  {title} {\bibinfo {title} {Exchange interactions and critical
  temperature of bulk and thin films of {M}n{S}i: A density functional theory
  study},\ }\href {https://doi.org/10.1103/PhysRevB.78.104402} {\bibfield
  {journal} {\bibinfo  {journal} {Phys. Rev. B}\ }\textbf {\bibinfo {volume}
  {78}},\ \bibinfo {pages} {104402} (\bibinfo {year} {2008})}\BibitemShut
  {NoStop}%
\bibitem [{\citenamefont {Fang}\ \emph {et~al.}(2022)\citenamefont {Fang},
  \citenamefont {Zhang}, \citenamefont {Wang}, \citenamefont {Yang},
  \citenamefont {Wu}, \citenamefont {Li}, \citenamefont {Xiao}, \citenamefont
  {Lin}, \citenamefont {Zheng}, \citenamefont {Li}, \citenamefont {Wang},
  \citenamefont {Rodolakis}, \citenamefont {Song}, \citenamefont {Wang},
  \citenamefont {Cao},\ and\ \citenamefont {Liu}}]{Fang22}%
  \BibitemOpen
  \bibfield  {author} {\bibinfo {author} {\bibfnamefont {Y.}~\bibnamefont
  {Fang}}, \bibinfo {author} {\bibfnamefont {H.}~\bibnamefont {Zhang}},
  \bibinfo {author} {\bibfnamefont {D.}~\bibnamefont {Wang}}, \bibinfo {author}
  {\bibfnamefont {G.}~\bibnamefont {Yang}}, \bibinfo {author} {\bibfnamefont
  {Y.}~\bibnamefont {Wu}}, \bibinfo {author} {\bibfnamefont {P.}~\bibnamefont
  {Li}}, \bibinfo {author} {\bibfnamefont {Z.}~\bibnamefont {Xiao}}, \bibinfo
  {author} {\bibfnamefont {T.}~\bibnamefont {Lin}}, \bibinfo {author}
  {\bibfnamefont {H.}~\bibnamefont {Zheng}}, \bibinfo {author} {\bibfnamefont
  {X.-L.}\ \bibnamefont {Li}}, \bibinfo {author} {\bibfnamefont {H.-H.}\
  \bibnamefont {Wang}}, \bibinfo {author} {\bibfnamefont {F.}~\bibnamefont
  {Rodolakis}}, \bibinfo {author} {\bibfnamefont {Y.}~\bibnamefont {Song}},
  \bibinfo {author} {\bibfnamefont {Y.}~\bibnamefont {Wang}}, \bibinfo {author}
  {\bibfnamefont {C.}~\bibnamefont {Cao}},\ and\ \bibinfo {author}
  {\bibfnamefont {Y.}~\bibnamefont {Liu}},\ }\bibfield  {title} {\bibinfo
  {title} {Quasiparticle characteristics of the weakly ferromagnetic
  \uppercase{H}und metal \uppercase{M}n\uppercase{S}i},\ }\href
  {https://doi.org/10.1103/PhysRevB.106.L161112} {\bibfield  {journal}
  {\bibinfo  {journal} {Phys. Rev. B}\ }\textbf {\bibinfo {volume} {106}},\
  \bibinfo {pages} {L161112} (\bibinfo {year} {2022})}\BibitemShut {NoStop}%
\bibitem [{\citenamefont {Dalmas~de R\'eotier}\ \emph
  {et~al.}(2024)\citenamefont {Dalmas~de R\'eotier}, \citenamefont {Yaouanc},
  \citenamefont {Lapertot}, \citenamefont {Wang}, \citenamefont {Amato},\ and\
  \citenamefont {Andreica}}]{Dalmas24}%
  \BibitemOpen
  \bibfield  {author} {\bibinfo {author} {\bibfnamefont {P.}~\bibnamefont
  {Dalmas~de R\'eotier}}, \bibinfo {author} {\bibfnamefont {A.}~\bibnamefont
  {Yaouanc}}, \bibinfo {author} {\bibfnamefont {G.}~\bibnamefont {Lapertot}},
  \bibinfo {author} {\bibfnamefont {C.}~\bibnamefont {Wang}}, \bibinfo {author}
  {\bibfnamefont {A.}~\bibnamefont {Amato}},\ and\ \bibinfo {author}
  {\bibfnamefont {D.}~\bibnamefont {Andreica}},\ }\bibfield  {title} {\bibinfo
  {title} {Experimental determination of the spin {H}amiltonian of the cubic
  chiral magnet {M}n{S}i},\ }\href
  {https://doi.org/10.1103/PhysRevB.109.L020408} {\bibfield  {journal}
  {\bibinfo  {journal} {Phys. Rev. B}\ }\textbf {\bibinfo {volume} {109}},\
  \bibinfo {pages} {L020408} (\bibinfo {year} {2024})}\BibitemShut {NoStop}%
\bibitem [{\citenamefont {Bor\'en}(1933)}]{Boren33}%
  \BibitemOpen
  \bibfield  {author} {\bibinfo {author} {\bibfnamefont {B.}~\bibnamefont
  {Bor\'en}},\ }\bibfield  {title} {\bibinfo {title} {Roentgenuntersuchung der
  legierungen von silicium mit chrom, mangan, kobalt und nickel},\ }\href@noop
  {} {\bibfield  {journal} {\bibinfo  {journal} {Ark. Kemi Mineral. Geol.}\
  }\textbf {\bibinfo {volume} {11A}},\ \bibinfo {pages} {1} (\bibinfo {year}
  {1933})}\BibitemShut {NoStop}%
\bibitem [{\citenamefont {Motoya}\ \emph {et~al.}(1976)\citenamefont {Motoya},
  \citenamefont {Yasuoka}, \citenamefont {Nakamura},\ and\ \citenamefont
  {Wernick}}]{Motoya76}%
  \BibitemOpen
  \bibfield  {author} {\bibinfo {author} {\bibfnamefont {K.}~\bibnamefont
  {Motoya}}, \bibinfo {author} {\bibfnamefont {H.}~\bibnamefont {Yasuoka}},
  \bibinfo {author} {\bibfnamefont {Y.}~\bibnamefont {Nakamura}},\ and\
  \bibinfo {author} {\bibfnamefont {J.~H.}\ \bibnamefont {Wernick}},\
  }\bibfield  {title} {\bibinfo {title} {Helical spin structure in
  \uppercase{M}n\uppercase{S}i-\uppercase{NMR} studies},\ }\href
  {https://doi.org/10.1016/0038-1098(76)90058-2} {\bibfield  {journal}
  {\bibinfo  {journal} {Solid State Commun.}\ }\textbf {\bibinfo {volume}
  {19}},\ \bibinfo {pages} {529} (\bibinfo {year} {1976})}\BibitemShut
  {NoStop}%
\bibitem [{\citenamefont {Ishikawa}\ \emph {et~al.}(1976)\citenamefont
  {Ishikawa}, \citenamefont {Tajima}, \citenamefont {Bloch},\ and\
  \citenamefont {Roth}}]{Ishikawa76}%
  \BibitemOpen
  \bibfield  {author} {\bibinfo {author} {\bibfnamefont {Y.}~\bibnamefont
  {Ishikawa}}, \bibinfo {author} {\bibfnamefont {K.}~\bibnamefont {Tajima}},
  \bibinfo {author} {\bibfnamefont {D.}~\bibnamefont {Bloch}},\ and\ \bibinfo
  {author} {\bibfnamefont {M.}~\bibnamefont {Roth}},\ }\bibfield  {title}
  {\bibinfo {title} {Helical spin structure in manganese silicide
  \uppercase{M}n\uppercase{S}i},\ }\href
  {https://doi.org/0.1016/0038-1098(76)90057-0} {\bibfield  {journal} {\bibinfo
   {journal} {Solid State Commun.}\ }\textbf {\bibinfo {volume} {19}},\
  \bibinfo {pages} {525} (\bibinfo {year} {1976})}\BibitemShut {NoStop}%
\bibitem [{\citenamefont {Date}\ \emph {et~al.}(1977)\citenamefont {Date},
  \citenamefont {Okuda},\ and\ \citenamefont {Kadowaki}}]{Date77}%
  \BibitemOpen
  \bibfield  {author} {\bibinfo {author} {\bibfnamefont {M.}~\bibnamefont
  {Date}}, \bibinfo {author} {\bibfnamefont {K.}~\bibnamefont {Okuda}},\ and\
  \bibinfo {author} {\bibfnamefont {K.}~\bibnamefont {Kadowaki}},\ }\bibfield
  {title} {\bibinfo {title} {Electron spin resonance in the itinerant-electron
  helical magnet \uppercase{M}n\uppercase{S}i},\ }\href
  {https://doi.org/10.1143/JPSJ.42.1555} {\bibfield  {journal} {\bibinfo
  {journal} {J. Phys. Soc. Jpn.}\ }\textbf {\bibinfo {volume} {42}},\ \bibinfo
  {pages} {1555} (\bibinfo {year} {1977})}\BibitemShut {NoStop}%
\bibitem [{\citenamefont {Stishov}\ \emph {et~al.}(2007)\citenamefont
  {Stishov}, \citenamefont {Petrova}, \citenamefont {Khasanov}, \citenamefont
  {Panova}, \citenamefont {Shikov}, \citenamefont {Lashley}, \citenamefont
  {Wu},\ and\ \citenamefont {Lograsso}}]{Stishov07}%
  \BibitemOpen
  \bibfield  {author} {\bibinfo {author} {\bibfnamefont {S.~M.}\ \bibnamefont
  {Stishov}}, \bibinfo {author} {\bibfnamefont {A.~E.}\ \bibnamefont
  {Petrova}}, \bibinfo {author} {\bibfnamefont {S.}~\bibnamefont {Khasanov}},
  \bibinfo {author} {\bibfnamefont {G.~K.}\ \bibnamefont {Panova}}, \bibinfo
  {author} {\bibfnamefont {A.~A.}\ \bibnamefont {Shikov}}, \bibinfo {author}
  {\bibfnamefont {J.~C.}\ \bibnamefont {Lashley}}, \bibinfo {author}
  {\bibfnamefont {D.}~\bibnamefont {Wu}},\ and\ \bibinfo {author}
  {\bibfnamefont {T.~A.}\ \bibnamefont {Lograsso}},\ }\bibfield  {title}
  {\bibinfo {title} {Magnetic phase transition in the itinerant helimagnet
  \uppercase{M}n\uppercase{S}i: Thermodynamic and transport properties},\
  }\href {https://doi.org/10.1103/PhysRevB.76.052405} {\bibfield  {journal}
  {\bibinfo  {journal} {Phys. Rev. B}\ }\textbf {\bibinfo {volume} {76}},\
  \bibinfo {pages} {052405} (\bibinfo {year} {2007})}\BibitemShut {NoStop}%
\bibitem [{\citenamefont {Pfleiderer}\ \emph {et~al.}(1997)\citenamefont
  {Pfleiderer}, \citenamefont {McMullan}, \citenamefont {Julian},\ and\
  \citenamefont {Lonzarich}}]{Pfleiderer97}%
  \BibitemOpen
  \bibfield  {author} {\bibinfo {author} {\bibfnamefont {C.}~\bibnamefont
  {Pfleiderer}}, \bibinfo {author} {\bibfnamefont {G.~J.}\ \bibnamefont
  {McMullan}}, \bibinfo {author} {\bibfnamefont {S.~R.}\ \bibnamefont
  {Julian}},\ and\ \bibinfo {author} {\bibfnamefont {G.~G.}\ \bibnamefont
  {Lonzarich}},\ }\bibfield  {title} {\bibinfo {title} {Magnetic quantum phase
  transition in \uppercase{M}n\uppercase{S}i under hydrostatic pressure},\
  }\href {https://doi.org/10.1103/PhysRevB.55.8330} {\bibfield  {journal}
  {\bibinfo  {journal} {Phys. Rev. B}\ }\textbf {\bibinfo {volume} {55}},\
  \bibinfo {pages} {8330} (\bibinfo {year} {1997})}\BibitemShut {NoStop}%
\bibitem [{\citenamefont {F{\aa}k}\ \emph {et~al.}(2005)\citenamefont
  {F{\aa}k}, \citenamefont {Sadykov}, \citenamefont {Flouquet},\ and\
  \citenamefont {Lapertot}}]{Fak05}%
  \BibitemOpen
  \bibfield  {author} {\bibinfo {author} {\bibfnamefont {B.}~\bibnamefont
  {F{\aa}k}}, \bibinfo {author} {\bibfnamefont {R.~A.}\ \bibnamefont
  {Sadykov}}, \bibinfo {author} {\bibfnamefont {J.}~\bibnamefont {Flouquet}},\
  and\ \bibinfo {author} {\bibfnamefont {G.}~\bibnamefont {Lapertot}},\
  }\bibfield  {title} {\bibinfo {title} {Pressure dependence of the magnetic
  structure of the itinerant electron magnet \uppercase{M}n\uppercase{S}i},\
  }\href {https://doi.org/10.1088/0953-8984/17/10/018} {\bibfield  {journal}
  {\bibinfo  {journal} {J. Phys.: Condens. Matter}\ }\textbf {\bibinfo {volume}
  {17}},\ \bibinfo {pages} {1635} (\bibinfo {year} {2005})}\BibitemShut
  {NoStop}%
\bibitem [{\citenamefont {Bannenberg}\ \emph {et~al.}(2019)\citenamefont
  {Bannenberg}, \citenamefont {Sadykov}, \citenamefont {Dalgliesh},
  \citenamefont {Goodway}, \citenamefont {Schlagel}, \citenamefont {Lograsso},
  \citenamefont {Falus}, \citenamefont {Leli\`evre-Berna}, \citenamefont
  {Leonov},\ and\ \citenamefont {Pappas}}]{Bannenberg19}%
  \BibitemOpen
  \bibfield  {author} {\bibinfo {author} {\bibfnamefont {L.~J.}\ \bibnamefont
  {Bannenberg}}, \bibinfo {author} {\bibfnamefont {R.}~\bibnamefont {Sadykov}},
  \bibinfo {author} {\bibfnamefont {R.~M.}\ \bibnamefont {Dalgliesh}}, \bibinfo
  {author} {\bibfnamefont {C.}~\bibnamefont {Goodway}}, \bibinfo {author}
  {\bibfnamefont {D.~L.}\ \bibnamefont {Schlagel}}, \bibinfo {author}
  {\bibfnamefont {T.~A.}\ \bibnamefont {Lograsso}}, \bibinfo {author}
  {\bibfnamefont {P.}~\bibnamefont {Falus}}, \bibinfo {author} {\bibfnamefont
  {E.}~\bibnamefont {Leli\`evre-Berna}}, \bibinfo {author} {\bibfnamefont
  {A.~O.}\ \bibnamefont {Leonov}},\ and\ \bibinfo {author} {\bibfnamefont
  {C.}~\bibnamefont {Pappas}},\ }\bibfield  {title} {\bibinfo {title}
  {Skyrmions and spirals in \uppercase{M}n\uppercase{S}i under hydrostatic
  pressure},\ }\href {https://doi.org/10.1103/PhysRevB.100.054447} {\bibfield
  {journal} {\bibinfo  {journal} {Phys. Rev. B}\ }\textbf {\bibinfo {volume}
  {100}},\ \bibinfo {pages} {054447} (\bibinfo {year} {2019})}\BibitemShut
  {NoStop}%
\bibitem [{\citenamefont {Andreica}\ \emph {et~al.}(2010)\citenamefont
  {Andreica}, \citenamefont {{Dalmas de R\'eotier}}, \citenamefont {Yaouanc},
  \citenamefont {Amato},\ and\ \citenamefont {Lapertot}}]{Andreica10}%
  \BibitemOpen
  \bibfield  {author} {\bibinfo {author} {\bibfnamefont {D.}~\bibnamefont
  {Andreica}}, \bibinfo {author} {\bibfnamefont {P.}~\bibnamefont {{Dalmas de
  R\'eotier}}}, \bibinfo {author} {\bibfnamefont {A.}~\bibnamefont {Yaouanc}},
  \bibinfo {author} {\bibfnamefont {A.}~\bibnamefont {Amato}},\ and\ \bibinfo
  {author} {\bibfnamefont {G.}~\bibnamefont {Lapertot}},\ }\bibfield  {title}
  {\bibinfo {title} {Absence of magnetic phase separation in
  \uppercase{M}n\uppercase{S}i under pressure},\ }\href
  {https://doi.org/10.1103/PhysRevB.81.060412} {\bibfield  {journal} {\bibinfo
  {journal} {Phys. Rev. B}\ }\textbf {\bibinfo {volume} {81}},\ \bibinfo
  {pages} {060412(R)} (\bibinfo {year} {2010})}\BibitemShut {NoStop}%
\bibitem [{SM()}]{SM}%
  \BibitemOpen
  \href@noop {} {}\bibinfo {note} {See Supplemental Material for information on
  the crystal structure and the determination of $T_c$, for further ZF-$\mu$SR
  spectra, for alternative models for the spectra recorded at 1.25~GPa, and for
  a discussion of $m(T)$ and of the scaling law; see also
  Refs.~\onlinecite{Takigawa80,Grigoriev06,Kadono90,Bak80,Chaikin95,Garst17,Kormann09}
  cited therein.}\BibitemShut {Stop}%
\bibitem [{\citenamefont {Brazhkin}\ \emph {et~al.}(2016)\citenamefont
  {Brazhkin}, \citenamefont {Dzhavadov},\ and\ \citenamefont
  {El’kin}}]{Brazhkin16}%
  \BibitemOpen
  \bibfield  {author} {\bibinfo {author} {\bibfnamefont {V.~V.}\ \bibnamefont
  {Brazhkin}}, \bibinfo {author} {\bibfnamefont {L.~N.}\ \bibnamefont
  {Dzhavadov}},\ and\ \bibinfo {author} {\bibfnamefont {F.~S.}\ \bibnamefont
  {El’kin}},\ }\bibfield  {title} {\bibinfo {title} {Study of the
  compressibility of {F}e{S}i, {M}n{S}i, and {C}o{S}$_2$ transition-metal
  compounds at high pressures},\ }\href
  {https://doi.org/10.1134/S0021364016140083} {\bibfield  {journal} {\bibinfo
  {journal} {JETP Lett.}\ }\textbf {\bibinfo {volume} {104}},\ \bibinfo {pages}
  {99} (\bibinfo {year} {2016})}\BibitemShut {NoStop}%
\bibitem [{\citenamefont {Dalmas~de R\'eotier}\ \emph
  {et~al.}(2016)\citenamefont {Dalmas~de R\'eotier}, \citenamefont
  {Maisuradze}, \citenamefont {Yaouanc}, \citenamefont {Roessli}, \citenamefont
  {Amato}, \citenamefont {Andreica},\ and\ \citenamefont
  {Lapertot}}]{Dalmas16}%
  \BibitemOpen
  \bibfield  {author} {\bibinfo {author} {\bibfnamefont {P.}~\bibnamefont
  {Dalmas~de R\'eotier}}, \bibinfo {author} {\bibfnamefont {A.}~\bibnamefont
  {Maisuradze}}, \bibinfo {author} {\bibfnamefont {A.}~\bibnamefont {Yaouanc}},
  \bibinfo {author} {\bibfnamefont {B.}~\bibnamefont {Roessli}}, \bibinfo
  {author} {\bibfnamefont {A.}~\bibnamefont {Amato}}, \bibinfo {author}
  {\bibfnamefont {D.}~\bibnamefont {Andreica}},\ and\ \bibinfo {author}
  {\bibfnamefont {G.}~\bibnamefont {Lapertot}},\ }\bibfield  {title} {\bibinfo
  {title} {Determination of the zero-field magnetic structure of the helimagnet
  \uppercase{M}n\uppercase{S}i at low temperature},\ }\href
  {https://doi.org/10.1103/PhysRevB.93.144419} {\bibfield  {journal} {\bibinfo
  {journal} {Phys. Rev. B}\ }\textbf {\bibinfo {volume} {93}},\ \bibinfo
  {pages} {144419} (\bibinfo {year} {2016})}\BibitemShut {NoStop}%
\bibitem [{\citenamefont {Huber}\ \emph {et~al.}(1975)\citenamefont {Huber},
  \citenamefont {Maple}, \citenamefont {Wohlleben},\ and\ \citenamefont
  {Knapp}}]{Huber75}%
  \BibitemOpen
  \bibfield  {author} {\bibinfo {author} {\bibfnamefont {J.}~\bibnamefont
  {Huber}}, \bibinfo {author} {\bibfnamefont {M.}~\bibnamefont {Maple}},
  \bibinfo {author} {\bibfnamefont {D.}~\bibnamefont {Wohlleben}},\ and\
  \bibinfo {author} {\bibfnamefont {G.}~\bibnamefont {Knapp}},\ }\bibfield
  {title} {\bibinfo {title} {Magnetic properties of {Z}r{Z}n$_2$ under
  pressure},\ }\href
  {https://doi.org/https://doi.org/10.1016/0038-1098(75)90577-3} {\bibfield
  {journal} {\bibinfo  {journal} {Solid State Commun.}\ }\textbf {\bibinfo
  {volume} {16}},\ \bibinfo {pages} {211} (\bibinfo {year} {1975})}\BibitemShut
  {NoStop}%
\bibitem [{\citenamefont {Mathur}\ \emph {et~al.}(1998)\citenamefont {Mathur},
  \citenamefont {Grosche}, \citenamefont {Julian}, \citenamefont {Walker},
  \citenamefont {Freye}, \citenamefont {Haselwimmer},\ and\ \citenamefont
  {Lonzarich}}]{Mathur98}%
  \BibitemOpen
  \bibfield  {author} {\bibinfo {author} {\bibfnamefont {N.~D.}\ \bibnamefont
  {Mathur}}, \bibinfo {author} {\bibfnamefont {F.~M.}\ \bibnamefont {Grosche}},
  \bibinfo {author} {\bibfnamefont {S.~R.}\ \bibnamefont {Julian}}, \bibinfo
  {author} {\bibfnamefont {I.~R.}\ \bibnamefont {Walker}}, \bibinfo {author}
  {\bibfnamefont {D.~M.}\ \bibnamefont {Freye}}, \bibinfo {author}
  {\bibfnamefont {R.~K.~W.}\ \bibnamefont {Haselwimmer}},\ and\ \bibinfo
  {author} {\bibfnamefont {G.~G.}\ \bibnamefont {Lonzarich}},\ }\bibfield
  {title} {\bibinfo {title} {Magnetically mediated superconductivity in heavy
  fermion compounds},\ }\href {https://doi.org/10.1038/27838} {\bibfield
  {journal} {\bibinfo  {journal} {Nature}\ }\textbf {\bibinfo {volume} {394}},\
  \bibinfo {pages} {39} (\bibinfo {year} {1998})}\BibitemShut {NoStop}%
\bibitem [{\citenamefont {Saxena}\ \emph {et~al.}(2000)\citenamefont {Saxena},
  \citenamefont {Agarwal}, \citenamefont {Ahilan}, \citenamefont {Grosche},
  \citenamefont {Haselwimmer}, \citenamefont {Steiner}, \citenamefont {Pugh},
  \citenamefont {Walker}, \citenamefont {Julian}, \citenamefont {Monthoux},
  \citenamefont {Lonzarich}, \citenamefont {Huxley}, \citenamefont {Sheikin},
  \citenamefont {Braithwaite},\ and\ \citenamefont {Flouquet}}]{Saxena00}%
  \BibitemOpen
  \bibfield  {author} {\bibinfo {author} {\bibfnamefont {S.~S.}\ \bibnamefont
  {Saxena}}, \bibinfo {author} {\bibfnamefont {P.}~\bibnamefont {Agarwal}},
  \bibinfo {author} {\bibfnamefont {K.}~\bibnamefont {Ahilan}}, \bibinfo
  {author} {\bibfnamefont {F.~M.}\ \bibnamefont {Grosche}}, \bibinfo {author}
  {\bibfnamefont {R.~K.}\ \bibnamefont {Haselwimmer}}, \bibinfo {author}
  {\bibfnamefont {M.~J.}\ \bibnamefont {Steiner}}, \bibinfo {author}
  {\bibfnamefont {E.}~\bibnamefont {Pugh}}, \bibinfo {author} {\bibfnamefont
  {I.~R.}\ \bibnamefont {Walker}}, \bibinfo {author} {\bibfnamefont {S.~R.}\
  \bibnamefont {Julian}}, \bibinfo {author} {\bibfnamefont {P.}~\bibnamefont
  {Monthoux}}, \bibinfo {author} {\bibfnamefont {G.~G.}\ \bibnamefont
  {Lonzarich}}, \bibinfo {author} {\bibfnamefont {A.}~\bibnamefont {Huxley}},
  \bibinfo {author} {\bibfnamefont {I.~I.}\ \bibnamefont {Sheikin}}, \bibinfo
  {author} {\bibfnamefont {D.}~\bibnamefont {Braithwaite}},\ and\ \bibinfo
  {author} {\bibfnamefont {J.}~\bibnamefont {Flouquet}},\ }\bibfield  {title}
  {\bibinfo {title} {Superconductivity on the border of itinerant-electron
  ferromagnetism in \uppercase{U}\uppercase{G}e$_2$},\ }\href
  {https://doi.org/10.1038/35020500} {\bibfield  {journal} {\bibinfo  {journal}
  {Nature}\ }\textbf {\bibinfo {volume} {406}},\ \bibinfo {pages} {587}
  (\bibinfo {year} {2000})}\BibitemShut {NoStop}%
\bibitem [{\citenamefont {L\"ohneysen}\ \emph {et~al.}(2007)\citenamefont
  {L\"ohneysen}, \citenamefont {Rosch}, \citenamefont {Vojta},\ and\
  \citenamefont {W\"olfle}}]{Lohneysen07}%
  \BibitemOpen
  \bibfield  {author} {\bibinfo {author} {\bibfnamefont {H.~v.}\ \bibnamefont
  {L\"ohneysen}}, \bibinfo {author} {\bibfnamefont {A.}~\bibnamefont {Rosch}},
  \bibinfo {author} {\bibfnamefont {M.}~\bibnamefont {Vojta}},\ and\ \bibinfo
  {author} {\bibfnamefont {P.}~\bibnamefont {W\"olfle}},\ }\bibfield  {title}
  {\bibinfo {title} {Fermi-liquid instabilities at magnetic quantum phase
  transitions},\ }\href {https://doi.org/10.1103/RevModPhys.79.1015} {\bibfield
   {journal} {\bibinfo  {journal} {Rev. Mod. Phys.}\ }\textbf {\bibinfo
  {volume} {79}},\ \bibinfo {pages} {1015} (\bibinfo {year}
  {2007})}\BibitemShut {NoStop}%
\bibitem [{\citenamefont {Kawano}\ \emph {et~al.}(1994)\citenamefont {Kawano},
  \citenamefont {Fernandez‐Baca},\ and\ \citenamefont {Nicklow}}]{Kawano94}%
  \BibitemOpen
  \bibfield  {author} {\bibinfo {author} {\bibfnamefont {S.}~\bibnamefont
  {Kawano}}, \bibinfo {author} {\bibfnamefont {J.~A.}\ \bibnamefont
  {Fernandez‐Baca}},\ and\ \bibinfo {author} {\bibfnamefont {R.~M.}\
  \bibnamefont {Nicklow}},\ }\bibfield  {title} {\bibinfo {title} {Magnons in
  ferromagnetic terbium under high pressure},\ }\href
  {https://doi.org/10.1063/1.355457} {\bibfield  {journal} {\bibinfo  {journal}
  {J. Appl. Phys.}\ }\textbf {\bibinfo {volume} {75}},\ \bibinfo {pages} {6060}
  (\bibinfo {year} {1994})}\BibitemShut {NoStop}%
\bibitem [{\citenamefont {Hayashida}\ \emph {et~al.}(2019)\citenamefont
  {Hayashida}, \citenamefont {Matsumoto}, \citenamefont {Hagihala},
  \citenamefont {Kurita}, \citenamefont {Tanaka}, \citenamefont {Itoh},
  \citenamefont {Hong}, \citenamefont {Soda}, \citenamefont {Uwatoko},\ and\
  \citenamefont {Masuda}}]{Hayashida19}%
  \BibitemOpen
  \bibfield  {author} {\bibinfo {author} {\bibfnamefont {S.}~\bibnamefont
  {Hayashida}}, \bibinfo {author} {\bibfnamefont {M.}~\bibnamefont
  {Matsumoto}}, \bibinfo {author} {\bibfnamefont {M.}~\bibnamefont {Hagihala}},
  \bibinfo {author} {\bibfnamefont {N.}~\bibnamefont {Kurita}}, \bibinfo
  {author} {\bibfnamefont {H.}~\bibnamefont {Tanaka}}, \bibinfo {author}
  {\bibfnamefont {S.}~\bibnamefont {Itoh}}, \bibinfo {author} {\bibfnamefont
  {T.}~\bibnamefont {Hong}}, \bibinfo {author} {\bibfnamefont {M.}~\bibnamefont
  {Soda}}, \bibinfo {author} {\bibfnamefont {Y.}~\bibnamefont {Uwatoko}},\ and\
  \bibinfo {author} {\bibfnamefont {T.}~\bibnamefont {Masuda}},\ }\bibfield
  {title} {\bibinfo {title} {Novel excitations near quantum criticality in
  geometrically frustrated antiferromagnet {C}s{F}e{C}l$_3$},\ }\href
  {https://doi.org/10.1126/sciadv.aaw5639} {\bibfield  {journal} {\bibinfo
  {journal} {Sci. Adv.}\ }\textbf {\bibinfo {volume} {5}},\ \bibinfo {pages}
  {eaaw5639} (\bibinfo {year} {2019})}\BibitemShut {NoStop}%
\bibitem [{\citenamefont {Pfleiderer}\ \emph {et~al.}(2007)\citenamefont
  {Pfleiderer}, \citenamefont {B\"oni}, \citenamefont {Keller}, \citenamefont
  {R\"o{\ss}ler},\ and\ \citenamefont {Rosch}}]{Pfleiderer07}%
  \BibitemOpen
  \bibfield  {author} {\bibinfo {author} {\bibfnamefont {C.}~\bibnamefont
  {Pfleiderer}}, \bibinfo {author} {\bibfnamefont {P.}~\bibnamefont {B\"oni}},
  \bibinfo {author} {\bibfnamefont {T.}~\bibnamefont {Keller}}, \bibinfo
  {author} {\bibfnamefont {U.~K.}\ \bibnamefont {R\"o{\ss}ler}},\ and\ \bibinfo
  {author} {\bibfnamefont {A.}~\bibnamefont {Rosch}},\ }\bibfield  {title}
  {\bibinfo {title} {Non-{F}ermi liquid metal without quantum criticality},\
  }\href {https://doi.org/10.1126/science.1142644} {\bibfield  {journal}
  {\bibinfo  {journal} {Science}\ }\textbf {\bibinfo {volume} {316}},\ \bibinfo
  {pages} {1871} (\bibinfo {year} {2007})}\BibitemShut {NoStop}%
\bibitem [{\citenamefont {Chizhikov}\ and\ \citenamefont
  {Dmitrienko}(2012)}]{Chizhikov12}%
  \BibitemOpen
  \bibfield  {author} {\bibinfo {author} {\bibfnamefont {V.~A.}\ \bibnamefont
  {Chizhikov}}\ and\ \bibinfo {author} {\bibfnamefont {V.~E.}\ \bibnamefont
  {Dmitrienko}},\ }\bibfield  {title} {\bibinfo {title} {Frustrated magnetic
  helices in \uppercase{M}n\uppercase{S}i-type crystals},\ }\href
  {https://doi.org/10.1103/PhysRevB.85.014421} {\bibfield  {journal} {\bibinfo
  {journal} {Phys. Rev. B}\ }\textbf {\bibinfo {volume} {85}},\ \bibinfo
  {pages} {014421} (\bibinfo {year} {2012})}\BibitemShut {NoStop}%
\bibitem [{\citenamefont {Maleyev}(2006)}]{Maleyev06}%
  \BibitemOpen
  \bibfield  {author} {\bibinfo {author} {\bibfnamefont {S.~V.}\ \bibnamefont
  {Maleyev}},\ }\bibfield  {title} {\bibinfo {title} {Cubic magnets with
  \uppercase{D}zyaloshinskii-\uppercase{M}oriya interaction at low
  temperature},\ }\href {https://doi.org/10.1103/PhysRevB.73.174402} {\bibfield
   {journal} {\bibinfo  {journal} {Phys. Rev. B}\ }\textbf {\bibinfo {volume}
  {73}},\ \bibinfo {pages} {174402} (\bibinfo {year} {2006})}\BibitemShut
  {NoStop}%
\bibitem [{\citenamefont {Belitz}\ \emph {et~al.}(2006)\citenamefont {Belitz},
  \citenamefont {Kirkpatrick},\ and\ \citenamefont {Rosch}}]{Belitz06}%
  \BibitemOpen
  \bibfield  {author} {\bibinfo {author} {\bibfnamefont {D.}~\bibnamefont
  {Belitz}}, \bibinfo {author} {\bibfnamefont {T.~R.}\ \bibnamefont
  {Kirkpatrick}},\ and\ \bibinfo {author} {\bibfnamefont {A.}~\bibnamefont
  {Rosch}},\ }\bibfield  {title} {\bibinfo {title} {Theory of helimagnons in
  itinerant quantum systems},\ }\href
  {https://doi.org/10.1103/PhysRevB.73.054431} {\bibfield  {journal} {\bibinfo
  {journal} {Phys. Rev. B}\ }\textbf {\bibinfo {volume} {73}},\ \bibinfo
  {pages} {054431} (\bibinfo {year} {2006})}\BibitemShut {NoStop}%
\bibitem [{\citenamefont {van Kranendonk}\ and\ \citenamefont {van
  Vleck}(1958)}]{vanKranendonk58}%
  \BibitemOpen
  \bibfield  {author} {\bibinfo {author} {\bibfnamefont {J.}~\bibnamefont {van
  Kranendonk}}\ and\ \bibinfo {author} {\bibfnamefont {J.~H.}\ \bibnamefont
  {van Vleck}},\ }\bibfield  {title} {\bibinfo {title} {Spin waves},\ }\href
  {https://doi.org/10.1103/RevModPhys.30.1} {\bibfield  {journal} {\bibinfo
  {journal} {Rev. Mod. Phys.}\ }\textbf {\bibinfo {volume} {30}},\ \bibinfo
  {pages} {1} (\bibinfo {year} {1958})}\BibitemShut {NoStop}%
\bibitem [{Note1()}]{Note1}%
  \BibitemOpen
  \bibinfo {note} {This sample was already used for the studies published in
  Refs.~\protect \rev@citealp {Andreica10,Amato14}.}\BibitemShut {Stop}%
\bibitem [{\citenamefont {Shermadini}\ \emph {et~al.}(2017)\citenamefont
  {Shermadini}, \citenamefont {Khasanov}, \citenamefont {Elender},
  \citenamefont {Simutis}, \citenamefont {Guguchia}, \citenamefont {Kamenev},\
  and\ \citenamefont {Amato}}]{Shermadini17}%
  \BibitemOpen
  \bibfield  {author} {\bibinfo {author} {\bibfnamefont {Z.}~\bibnamefont
  {Shermadini}}, \bibinfo {author} {\bibfnamefont {R.}~\bibnamefont
  {Khasanov}}, \bibinfo {author} {\bibfnamefont {M.}~\bibnamefont {Elender}},
  \bibinfo {author} {\bibfnamefont {G.}~\bibnamefont {Simutis}}, \bibinfo
  {author} {\bibfnamefont {Z.}~\bibnamefont {Guguchia}}, \bibinfo {author}
  {\bibfnamefont {K.~V.}\ \bibnamefont {Kamenev}},\ and\ \bibinfo {author}
  {\bibfnamefont {A.}~\bibnamefont {Amato}},\ }\bibfield  {title} {\bibinfo
  {title} {A low-background piston–cylinder-type hybrid high pressure cell
  for muon-spin rotation/relaxation experiments},\ }\href
  {https://doi.org/10.1080/08957959.2017.1373773} {\bibfield  {journal}
  {\bibinfo  {journal} {High Press. Res.}\ }\textbf {\bibinfo {volume} {37}},\
  \bibinfo {pages} {449} (\bibinfo {year} {2017})}\BibitemShut {NoStop}%
\bibitem [{\citenamefont {Khasanov}\ \emph {et~al.}(2016)\citenamefont
  {Khasanov}, \citenamefont {Guguchia}, \citenamefont {Maisuradze},
  \citenamefont {Andreica}, \citenamefont {Elender}, \citenamefont {Raselli},
  \citenamefont {Shermadini}, \citenamefont {Goko}, \citenamefont {Knecht},
  \citenamefont {Morenzoni},\ and\ \citenamefont {Amato}}]{Khasanov16}%
  \BibitemOpen
  \bibfield  {author} {\bibinfo {author} {\bibfnamefont {R.}~\bibnamefont
  {Khasanov}}, \bibinfo {author} {\bibfnamefont {Z.}~\bibnamefont {Guguchia}},
  \bibinfo {author} {\bibfnamefont {A.}~\bibnamefont {Maisuradze}}, \bibinfo
  {author} {\bibfnamefont {D.}~\bibnamefont {Andreica}}, \bibinfo {author}
  {\bibfnamefont {M.}~\bibnamefont {Elender}}, \bibinfo {author} {\bibfnamefont
  {A.}~\bibnamefont {Raselli}}, \bibinfo {author} {\bibfnamefont
  {Z.}~\bibnamefont {Shermadini}}, \bibinfo {author} {\bibfnamefont
  {T.}~\bibnamefont {Goko}}, \bibinfo {author} {\bibfnamefont {F.}~\bibnamefont
  {Knecht}}, \bibinfo {author} {\bibfnamefont {E.}~\bibnamefont {Morenzoni}},\
  and\ \bibinfo {author} {\bibfnamefont {A.}~\bibnamefont {Amato}},\ }\bibfield
   {title} {\bibinfo {title} {High pressure research using muons at the
  \uppercase{P}aul \uppercase{S}cherrer \uppercase{I}nstitute},\ }\href
  {https://doi.org/10.1080/08957959.2016.1173690} {\bibfield  {journal}
  {\bibinfo  {journal} {High Press. Res.}\ }\textbf {\bibinfo {volume} {36}},\
  \bibinfo {pages} {140} (\bibinfo {year} {2016})}\BibitemShut {NoStop}%
\bibitem [{\citenamefont {Khasanov}(2022)}]{Khasanov22}%
  \BibitemOpen
  \bibfield  {author} {\bibinfo {author} {\bibfnamefont {R.}~\bibnamefont
  {Khasanov}},\ }\bibfield  {title} {\bibinfo {title} {Perspective on muon-spin
  rotation/relaxation under hydrostatic pressure},\ }\href
  {https://doi.org/10.1063/5.0119840} {\bibfield  {journal} {\bibinfo
  {journal} {J. Appl. Phys.}\ }\textbf {\bibinfo {volume} {132}},\ \bibinfo
  {pages} {190903} (\bibinfo {year} {2022})}\BibitemShut {NoStop}%
\bibitem [{\citenamefont {Yaouanc}\ and\ \citenamefont {{Dalmas de
  R\'eotier}}(2011)}]{Yaouanc11}%
  \BibitemOpen
  \bibfield  {author} {\bibinfo {author} {\bibfnamefont {A.}~\bibnamefont
  {Yaouanc}}\ and\ \bibinfo {author} {\bibfnamefont {P.}~\bibnamefont {{Dalmas
  de R\'eotier}}},\ }\href@noop {} {\emph {\bibinfo {title} {Muon Spin
  Rotation, Relaxation, and Resonance: Applications to Condensed Matter}}}\
  (\bibinfo  {publisher} {Oxford University Press},\ \bibinfo {address}
  {Oxford},\ \bibinfo {year} {2011})\BibitemShut {NoStop}%
\bibitem [{\citenamefont {Blundell}\ \emph {et~al.}(2022)\citenamefont
  {Blundell}, \citenamefont {{De Renzi}}, \citenamefont {Lancaster},\ and\
  \citenamefont {Pratt}}]{Blundell22}%
  \BibitemOpen
  \bibinfo {editor} {\bibfnamefont {S.~J.}\ \bibnamefont {Blundell}}, \bibinfo
  {editor} {\bibfnamefont {R.}~\bibnamefont {{De Renzi}}}, \bibinfo {editor}
  {\bibfnamefont {T.}~\bibnamefont {Lancaster}},\ and\ \bibinfo {editor}
  {\bibfnamefont {F.~L.}\ \bibnamefont {Pratt}},\ eds.,\ \href@noop {} {\emph
  {\bibinfo {title} {Muon Spectroscopy, an Introduction}}}\ (\bibinfo
  {publisher} {Oxford University Press},\ \bibinfo {address} {Oxford},\
  \bibinfo {year} {2022})\BibitemShut {NoStop}%
\bibitem [{\citenamefont {Amato}\ and\ \citenamefont
  {Morenzoni}(2024)}]{Amato24}%
  \BibitemOpen
  \bibfield  {author} {\bibinfo {author} {\bibfnamefont {A.}~\bibnamefont
  {Amato}}\ and\ \bibinfo {author} {\bibfnamefont {E.}~\bibnamefont
  {Morenzoni}},\ }\href@noop {} {\emph {\bibinfo {title} {Introduction to muon
  spin spectroscopy}}},\ Lecture notes in physics\ (\bibinfo  {publisher}
  {Springer},\ \bibinfo {year} {2024})\BibitemShut {NoStop}%
\bibitem [{\citenamefont {Amato}\ \emph {et~al.}(2014)\citenamefont {Amato},
  \citenamefont {{Dalmas de R\'eotier}}, \citenamefont {Andreica},
  \citenamefont {Yaouanc}, \citenamefont {Suter}, \citenamefont {Lapertot},
  \citenamefont {Pop}, \citenamefont {Morenzoni}, \citenamefont {Bonf\`a},
  \citenamefont {Bernardini},\ and\ \citenamefont {De~Renzi}}]{Amato14}%
  \BibitemOpen
  \bibfield  {author} {\bibinfo {author} {\bibfnamefont {A.}~\bibnamefont
  {Amato}}, \bibinfo {author} {\bibfnamefont {P.}~\bibnamefont {{Dalmas de
  R\'eotier}}}, \bibinfo {author} {\bibfnamefont {D.}~\bibnamefont {Andreica}},
  \bibinfo {author} {\bibfnamefont {A.}~\bibnamefont {Yaouanc}}, \bibinfo
  {author} {\bibfnamefont {A.}~\bibnamefont {Suter}}, \bibinfo {author}
  {\bibfnamefont {G.}~\bibnamefont {Lapertot}}, \bibinfo {author}
  {\bibfnamefont {I.~M.}\ \bibnamefont {Pop}}, \bibinfo {author} {\bibfnamefont
  {E.}~\bibnamefont {Morenzoni}}, \bibinfo {author} {\bibfnamefont
  {P.}~\bibnamefont {Bonf\`a}}, \bibinfo {author} {\bibfnamefont
  {F.}~\bibnamefont {Bernardini}},\ and\ \bibinfo {author} {\bibfnamefont
  {R.}~\bibnamefont {De~Renzi}},\ }\bibfield  {title} {\bibinfo {title}
  {Understanding the $\mu$\uppercase{SR} spectra of
  \uppercase{M}n\uppercase{S}i without magnetic polarons},\ }\href
  {https://doi.org/10.1103/PhysRevB.89.184425} {\bibfield  {journal} {\bibinfo
  {journal} {Phys. Rev. B}\ }\textbf {\bibinfo {volume} {89}},\ \bibinfo
  {pages} {184425} (\bibinfo {year} {2014})}\BibitemShut {NoStop}%
\bibitem [{\citenamefont {{Dalmas de R\'eotier}}\ \emph
  {et~al.}(2018)\citenamefont {{Dalmas de R\'eotier}}, \citenamefont {Yaouanc},
  \citenamefont {Amato}, \citenamefont {Maisuradze}, \citenamefont {Andreica},
  \citenamefont {Roessli}, \citenamefont {Goko}, \citenamefont {Scheuermann},\
  and\ \citenamefont {Lapertot}}]{Dalmas18}%
  \BibitemOpen
  \bibfield  {author} {\bibinfo {author} {\bibfnamefont {P.}~\bibnamefont
  {{Dalmas de R\'eotier}}}, \bibinfo {author} {\bibfnamefont {A.}~\bibnamefont
  {Yaouanc}}, \bibinfo {author} {\bibfnamefont {A.}~\bibnamefont {Amato}},
  \bibinfo {author} {\bibfnamefont {A.}~\bibnamefont {Maisuradze}}, \bibinfo
  {author} {\bibfnamefont {D.}~\bibnamefont {Andreica}}, \bibinfo {author}
  {\bibfnamefont {B.}~\bibnamefont {Roessli}}, \bibinfo {author} {\bibfnamefont
  {T.}~\bibnamefont {Goko}}, \bibinfo {author} {\bibfnamefont {R.}~\bibnamefont
  {Scheuermann}},\ and\ \bibinfo {author} {\bibfnamefont {G.}~\bibnamefont
  {Lapertot}},\ }\bibfield  {title} {\bibinfo {title} {On the robustness of the
  \uppercase{M}n\uppercase{S}i magnetic structure determined by muon spin
  rotation},\ }\href {https://doi.org/10.3390/qubs2030019} {\bibfield
  {journal} {\bibinfo  {journal} {Quantum Beam Sci.}\ }\textbf {\bibinfo
  {volume} {2}},\ \bibinfo {pages} {19} (\bibinfo {year} {2018})}\BibitemShut
  {NoStop}%
\bibitem [{\citenamefont {Thessieu}\ \emph {et~al.}(1999)\citenamefont
  {Thessieu}, \citenamefont {Kitaoka},\ and\ \citenamefont
  {Asayama}}]{Thessieu99}%
  \BibitemOpen
  \bibfield  {author} {\bibinfo {author} {\bibfnamefont {C.}~\bibnamefont
  {Thessieu}}, \bibinfo {author} {\bibfnamefont {Y.}~\bibnamefont {Kitaoka}},\
  and\ \bibinfo {author} {\bibfnamefont {K.}~\bibnamefont {Asayama}},\
  }\bibfield  {title} {\bibinfo {title} {Magnetic quantum phase transition in
  \uppercase{M}n\uppercase{S}i},\ }\href
  {https://doi.org/https://doi.org/10.1016/S0921-4526(98)01139-9} {\bibfield
  {journal} {\bibinfo  {journal} {Physica B}\ }\textbf {\bibinfo {volume}
  {259-261}},\ \bibinfo {pages} {847} (\bibinfo {year} {1999})}\BibitemShut
  {NoStop}%
\bibitem [{\citenamefont {Yu}\ \emph {et~al.}(2004)\citenamefont {Yu},
  \citenamefont {Zamborszky}, \citenamefont {Thompson}, \citenamefont {Sarrao},
  \citenamefont {Torelli}, \citenamefont {Fisk},\ and\ \citenamefont
  {Brown}}]{Yu04}%
  \BibitemOpen
  \bibfield  {author} {\bibinfo {author} {\bibfnamefont {W.}~\bibnamefont
  {Yu}}, \bibinfo {author} {\bibfnamefont {F.}~\bibnamefont {Zamborszky}},
  \bibinfo {author} {\bibfnamefont {J.~D.}\ \bibnamefont {Thompson}}, \bibinfo
  {author} {\bibfnamefont {J.~L.}\ \bibnamefont {Sarrao}}, \bibinfo {author}
  {\bibfnamefont {M.~E.}\ \bibnamefont {Torelli}}, \bibinfo {author}
  {\bibfnamefont {Z.}~\bibnamefont {Fisk}},\ and\ \bibinfo {author}
  {\bibfnamefont {S.~E.}\ \bibnamefont {Brown}},\ }\bibfield  {title} {\bibinfo
  {title} {Phase inhomogeneity of the itinerant ferromagnet
  \uppercase{M}n\uppercase{S}i at high pressures},\ }\href
  {https://doi.org/10.1103/PhysRevLett.92.086403} {\bibfield  {journal}
  {\bibinfo  {journal} {Phys. Rev. Lett.}\ }\textbf {\bibinfo {volume} {92}},\
  \bibinfo {pages} {086403} (\bibinfo {year} {2004})}\BibitemShut {NoStop}%
\bibitem [{Note2()}]{Note2}%
  \BibitemOpen
  \bibinfo {note} {The exchange $J$ is deduced from the spin wave dispersion
  $\protect \hbar \omega ({\protect \bf q}) = D_{\protect \rm sw}q^2$ valid for
  a ferromagnet, with $D_{\protect \rm sw} = JSa_{\protect \rm latt}^2$. The
  value for $J$ quoted in the main text is obtained for $D_{\protect \rm sw}$ =
  21\protect \tmspace +\thinmuskip {.1667em}(1)\protect \tmspace +\thinmuskip
  {.1667em}meV\protect \tmspace +\thinmuskip {.1667em}\r A$^2$ (Ref.~\protect
  \rev@citealp {Sato16}, see also Refs.~\protect \rev@citealp
  {Semadeni99,Boni11}) increased by $\approx 20\%$ to account for the fact that
  the measurements were performed at 27~K rather than at low temperature \cite
  {Ishikawa77}.}\BibitemShut {Stop}%
\bibitem [{\citenamefont {Sato}\ \emph {et~al.}(2016)\citenamefont {Sato},
  \citenamefont {Okuyama}, \citenamefont {Hong}, \citenamefont {Kikkawa},
  \citenamefont {Taguchi}, \citenamefont {Arima},\ and\ \citenamefont
  {Tokura}}]{Sato16}%
  \BibitemOpen
  \bibfield  {author} {\bibinfo {author} {\bibfnamefont {T.~J.}\ \bibnamefont
  {Sato}}, \bibinfo {author} {\bibfnamefont {D.}~\bibnamefont {Okuyama}},
  \bibinfo {author} {\bibfnamefont {T.}~\bibnamefont {Hong}}, \bibinfo {author}
  {\bibfnamefont {A.}~\bibnamefont {Kikkawa}}, \bibinfo {author} {\bibfnamefont
  {Y.}~\bibnamefont {Taguchi}}, \bibinfo {author} {\bibfnamefont {T.-h.}\
  \bibnamefont {Arima}},\ and\ \bibinfo {author} {\bibfnamefont
  {Y.}~\bibnamefont {Tokura}},\ }\bibfield  {title} {\bibinfo {title} {Magnon
  dispersion shift in the induced ferromagnetic phase of noncentrosymmetric
  \uppercase{M}n\uppercase{S}i},\ }\href
  {https://doi.org/10.1103/PhysRevB.94.144420} {\bibfield  {journal} {\bibinfo
  {journal} {Phys. Rev. B}\ }\textbf {\bibinfo {volume} {94}},\ \bibinfo
  {pages} {144420} (\bibinfo {year} {2016})}\BibitemShut {NoStop}%
\bibitem [{\citenamefont {B\"oni}\ \emph {et~al.}(2011)\citenamefont {B\"oni},
  \citenamefont {Roessli},\ and\ \citenamefont {Hradil}}]{Boni11}%
  \BibitemOpen
  \bibfield  {author} {\bibinfo {author} {\bibfnamefont {P.}~\bibnamefont
  {B\"oni}}, \bibinfo {author} {\bibfnamefont {B.}~\bibnamefont {Roessli}},\
  and\ \bibinfo {author} {\bibfnamefont {K.}~\bibnamefont {Hradil}},\
  }\bibfield  {title} {\bibinfo {title} {Inelastic neutron and x-ray scattering
  from incommensurate magnetic systems},\ }\href
  {https://doi.org/10.1088/0953-8984/23/25/254209} {\bibfield  {journal}
  {\bibinfo  {journal} {J. Phys.: Condens. Matter}\ }\textbf {\bibinfo {volume}
  {23}},\ \bibinfo {pages} {254209} (\bibinfo {year} {2011})}\BibitemShut
  {NoStop}%
\bibitem [{\citenamefont {Semadeni}\ \emph {et~al.}(1999)\citenamefont
  {Semadeni}, \citenamefont {B\"oni}, \citenamefont {Endoh}, \citenamefont
  {Roessli},\ and\ \citenamefont {Shirane}}]{Semadeni99}%
  \BibitemOpen
  \bibfield  {author} {\bibinfo {author} {\bibfnamefont {F.}~\bibnamefont
  {Semadeni}}, \bibinfo {author} {\bibfnamefont {P.}~\bibnamefont {B\"oni}},
  \bibinfo {author} {\bibfnamefont {Y.}~\bibnamefont {Endoh}}, \bibinfo
  {author} {\bibfnamefont {B.}~\bibnamefont {Roessli}},\ and\ \bibinfo {author}
  {\bibfnamefont {G.}~\bibnamefont {Shirane}},\ }\bibfield  {title} {\bibinfo
  {title} {Direct observation of spin-flip excitations in
  \uppercase{M}n\uppercase{S}i},\ }\href
  {https://doi.org/10.1016/S0921-4526(99)00077-0} {\bibfield  {journal}
  {\bibinfo  {journal} {Physica B}\ }\textbf {\bibinfo {volume} {267-268}},\
  \bibinfo {pages} {248} (\bibinfo {year} {1999})}\BibitemShut {NoStop}%
\bibitem [{Note3()}]{Note3}%
  \BibitemOpen
  \bibinfo {note} {We should note that contrasting values are found in the
  literature, e.g.\ $J$ = 11.4 or 21.8~meV \cite {Chen20,Jin23}. However they
  arise from studies focused at relatively large scattering vectors and energy
  transfers where the neutron intensities associated with the spin wave
  excitations and Stoner continuum interfere: this could explain the
  discrepancy with Refs.~\protect \rev@citealp
  {Sato16,Boni11,Semadeni99}.}\BibitemShut {Stop}%
\bibitem [{\citenamefont {Wang}\ \emph {et~al.}(1982)\citenamefont {Wang},
  \citenamefont {Prange},\ and\ \citenamefont {Korenman}}]{Wang82}%
  \BibitemOpen
  \bibfield  {author} {\bibinfo {author} {\bibfnamefont {C.~S.}\ \bibnamefont
  {Wang}}, \bibinfo {author} {\bibfnamefont {R.~E.}\ \bibnamefont {Prange}},\
  and\ \bibinfo {author} {\bibfnamefont {V.}~\bibnamefont {Korenman}},\
  }\bibfield  {title} {\bibinfo {title} {Magnetism in iron and nickel},\ }\href
  {https://doi.org/10.1103/PhysRevB.25.5766} {\bibfield  {journal} {\bibinfo
  {journal} {Phys. Rev. B}\ }\textbf {\bibinfo {volume} {25}},\ \bibinfo
  {pages} {5766} (\bibinfo {year} {1982})}\BibitemShut {NoStop}%
\bibitem [{\citenamefont {Bogolyubov}\ and\ \citenamefont
  {Tyablikov}(1959)}]{Bogolyubov59}%
  \BibitemOpen
  \bibfield  {author} {\bibinfo {author} {\bibfnamefont {N.~N.}\ \bibnamefont
  {Bogolyubov}}\ and\ \bibinfo {author} {\bibfnamefont {S.~V.}\ \bibnamefont
  {Tyablikov}},\ }\bibfield  {title} {\bibinfo {title} {Retarded and advanced
  {G}reen functions in statistical physics},\ }\href@noop {} {\bibfield
  {journal} {\bibinfo  {journal} {Sov. Phys. Dokl.}\ }\textbf {\bibinfo
  {volume} {4}},\ \bibinfo {pages} {589} (\bibinfo {year} {1959})}\BibitemShut
  {NoStop}%
\bibitem [{\citenamefont {Povzner}\ \emph {et~al.}(2016)\citenamefont
  {Povzner}, \citenamefont {Volkov},\ and\ \citenamefont
  {Yasyulevich}}]{Povzner16}%
  \BibitemOpen
  \bibfield  {author} {\bibinfo {author} {\bibfnamefont {A.~A.}\ \bibnamefont
  {Povzner}}, \bibinfo {author} {\bibfnamefont {A.~G.}\ \bibnamefont
  {Volkov}},\ and\ \bibinfo {author} {\bibfnamefont {I.~A.}\ \bibnamefont
  {Yasyulevich}},\ }\bibfield  {title} {\bibinfo {title} {Effect of pressure on
  helical ferromagnetism in {M}n{S}i},\ }\href
  {https://doi.org/10.1007/s11182-016-0866-9} {\bibfield  {journal} {\bibinfo
  {journal} {Russ. Phys. J.}\ }\textbf {\bibinfo {volume} {59}},\ \bibinfo
  {pages} {1002} (\bibinfo {year} {2016})}\BibitemShut {NoStop}%
\bibitem [{\citenamefont {Petrova}\ and\ \citenamefont
  {Stishov}(2012)}]{Petrova12}%
  \BibitemOpen
  \bibfield  {author} {\bibinfo {author} {\bibfnamefont {A.~E.}\ \bibnamefont
  {Petrova}}\ and\ \bibinfo {author} {\bibfnamefont {S.~M.}\ \bibnamefont
  {Stishov}},\ }\bibfield  {title} {\bibinfo {title} {Phase diagram of the
  itinerant helimagnet {M}n{S}i from high-pressure resistivity measurements and
  the quantum criticality problem},\ }\href
  {https://doi.org/10.1103/PhysRevB.86.174407} {\bibfield  {journal} {\bibinfo
  {journal} {Phys. Rev. B}\ }\textbf {\bibinfo {volume} {86}},\ \bibinfo
  {pages} {174407} (\bibinfo {year} {2012})}\BibitemShut {NoStop}%
\bibitem [{\citenamefont {Pappas}\ \emph {et~al.}(2021)\citenamefont {Pappas},
  \citenamefont {Leonov}, \citenamefont {Bannenberg}, \citenamefont {Fouquet},
  \citenamefont {Wolf},\ and\ \citenamefont {Weber}}]{Pappas21}%
  \BibitemOpen
  \bibfield  {author} {\bibinfo {author} {\bibfnamefont {C.}~\bibnamefont
  {Pappas}}, \bibinfo {author} {\bibfnamefont {A.~O.}\ \bibnamefont {Leonov}},
  \bibinfo {author} {\bibfnamefont {L.~J.}\ \bibnamefont {Bannenberg}},
  \bibinfo {author} {\bibfnamefont {P.}~\bibnamefont {Fouquet}}, \bibinfo
  {author} {\bibfnamefont {T.}~\bibnamefont {Wolf}},\ and\ \bibinfo {author}
  {\bibfnamefont {F.}~\bibnamefont {Weber}},\ }\bibfield  {title} {\bibinfo
  {title} {Evolution of helimagnetic correlations when approaching the quantum
  critical point of {M}n$_{1-x}${F}e$_x${S}i},\ }\href
  {https://doi.org/10.1103/PhysRevResearch.3.013019} {\bibfield  {journal}
  {\bibinfo  {journal} {Phys. Rev. Res.}\ }\textbf {\bibinfo {volume} {3}},\
  \bibinfo {pages} {013019} (\bibinfo {year} {2021})}\BibitemShut {NoStop}%
\bibitem [{\citenamefont {Pedrazzini}\ \emph {et~al.}(2007)\citenamefont
  {Pedrazzini}, \citenamefont {Wilhelm}, \citenamefont {Jaccard}, \citenamefont
  {Jarlborg}, \citenamefont {Schmidt}, \citenamefont {Hanfland}, \citenamefont
  {Akselrud}, \citenamefont {Yuan}, \citenamefont {Schwarz}, \citenamefont
  {Grin},\ and\ \citenamefont {Steglich}}]{Pedrazzini07}%
  \BibitemOpen
  \bibfield  {author} {\bibinfo {author} {\bibfnamefont {P.}~\bibnamefont
  {Pedrazzini}}, \bibinfo {author} {\bibfnamefont {H.}~\bibnamefont {Wilhelm}},
  \bibinfo {author} {\bibfnamefont {D.}~\bibnamefont {Jaccard}}, \bibinfo
  {author} {\bibfnamefont {T.}~\bibnamefont {Jarlborg}}, \bibinfo {author}
  {\bibfnamefont {M.}~\bibnamefont {Schmidt}}, \bibinfo {author} {\bibfnamefont
  {M.}~\bibnamefont {Hanfland}}, \bibinfo {author} {\bibfnamefont
  {L.}~\bibnamefont {Akselrud}}, \bibinfo {author} {\bibfnamefont {H.~Q.}\
  \bibnamefont {Yuan}}, \bibinfo {author} {\bibfnamefont {U.}~\bibnamefont
  {Schwarz}}, \bibinfo {author} {\bibfnamefont {Y.}~\bibnamefont {Grin}},\ and\
  \bibinfo {author} {\bibfnamefont {F.}~\bibnamefont {Steglich}},\ }\bibfield
  {title} {\bibinfo {title} {Metallic state in cubic {F}e{G}e beyond its
  quantum phase transition},\ }\href
  {https://doi.org/10.1103/PhysRevLett.98.047204} {\bibfield  {journal}
  {\bibinfo  {journal} {Phys. Rev. Lett.}\ }\textbf {\bibinfo {volume} {98}},\
  \bibinfo {pages} {047204} (\bibinfo {year} {2007})}\BibitemShut {NoStop}%
\bibitem [{\citenamefont {Deutsch}\ \emph {et~al.}(2014)\citenamefont
  {Deutsch}, \citenamefont {Makarova}, \citenamefont {Hansen}, \citenamefont
  {Fernandez-Diaz}, \citenamefont {Sidorov}, \citenamefont {Tsvyashchenko},
  \citenamefont {Fomicheva}, \citenamefont {Porcher}, \citenamefont {Petit},
  \citenamefont {Koepernik}, \citenamefont {R\"o\ss{}ler},\ and\ \citenamefont
  {Mirebeau}}]{Deutsch14}%
  \BibitemOpen
  \bibfield  {author} {\bibinfo {author} {\bibfnamefont {M.}~\bibnamefont
  {Deutsch}}, \bibinfo {author} {\bibfnamefont {O.~L.}\ \bibnamefont
  {Makarova}}, \bibinfo {author} {\bibfnamefont {T.~C.}\ \bibnamefont
  {Hansen}}, \bibinfo {author} {\bibfnamefont {M.~T.}\ \bibnamefont
  {Fernandez-Diaz}}, \bibinfo {author} {\bibfnamefont {V.~A.}\ \bibnamefont
  {Sidorov}}, \bibinfo {author} {\bibfnamefont {A.~V.}\ \bibnamefont
  {Tsvyashchenko}}, \bibinfo {author} {\bibfnamefont {L.~N.}\ \bibnamefont
  {Fomicheva}}, \bibinfo {author} {\bibfnamefont {F.}~\bibnamefont {Porcher}},
  \bibinfo {author} {\bibfnamefont {S.}~\bibnamefont {Petit}}, \bibinfo
  {author} {\bibfnamefont {K.}~\bibnamefont {Koepernik}}, \bibinfo {author}
  {\bibfnamefont {U.~K.}\ \bibnamefont {R\"o\ss{}ler}},\ and\ \bibinfo {author}
  {\bibfnamefont {I.}~\bibnamefont {Mirebeau}},\ }\bibfield  {title} {\bibinfo
  {title} {Two-step pressure-induced collapse of magnetic order in the
  \uppercase{M}n\uppercase{G}e chiral magnet},\ }\href
  {https://doi.org/10.1103/PhysRevB.89.180407} {\bibfield  {journal} {\bibinfo
  {journal} {Phys. Rev. B}\ }\textbf {\bibinfo {volume} {89}},\ \bibinfo
  {pages} {180407(R)} (\bibinfo {year} {2014})}\BibitemShut {NoStop}%
\bibitem [{\citenamefont {Ruderman}\ and\ \citenamefont
  {Kittel}(1954)}]{Ruderman54}%
  \BibitemOpen
  \bibfield  {author} {\bibinfo {author} {\bibfnamefont {M.~A.}\ \bibnamefont
  {Ruderman}}\ and\ \bibinfo {author} {\bibfnamefont {C.}~\bibnamefont
  {Kittel}},\ }\bibfield  {title} {\bibinfo {title} {Indirect exchange coupling
  of nuclear magnetic moments by conduction electrons},\ }\href
  {https://doi.org/10.1103/PhysRev.96.99} {\bibfield  {journal} {\bibinfo
  {journal} {Phys. Rev.}\ }\textbf {\bibinfo {volume} {96}},\ \bibinfo {pages}
  {99} (\bibinfo {year} {1954})}\BibitemShut {NoStop}%
\bibitem [{\citenamefont {Kasuya}(1956)}]{Kasuya56}%
  \BibitemOpen
  \bibfield  {author} {\bibinfo {author} {\bibfnamefont {T.}~\bibnamefont
  {Kasuya}},\ }\bibfield  {title} {\bibinfo {title} {A theory of metallic
  ferro- and antiferromagnetism on {Z}ener's model},\ }\href
  {https://doi.org/10.1143/PTP.16.45} {\bibfield  {journal} {\bibinfo
  {journal} {Prog. Theor. Phys.}\ }\textbf {\bibinfo {volume} {16}},\ \bibinfo
  {pages} {45} (\bibinfo {year} {1956})}\BibitemShut {NoStop}%
\bibitem [{\citenamefont {Yosida}(1957)}]{Yosida57}%
  \BibitemOpen
  \bibfield  {author} {\bibinfo {author} {\bibfnamefont {K.}~\bibnamefont
  {Yosida}},\ }\bibfield  {title} {\bibinfo {title} {Magnetic properties of
  {C}u-{M}n alloys},\ }\href {https://doi.org/10.1103/PhysRev.106.893}
  {\bibfield  {journal} {\bibinfo  {journal} {Phys. Rev.}\ }\textbf {\bibinfo
  {volume} {106}},\ \bibinfo {pages} {893} (\bibinfo {year}
  {1957})}\BibitemShut {NoStop}%
\bibitem [{\citenamefont {Pfleiderer}\ \emph {et~al.}(2001)\citenamefont
  {Pfleiderer}, \citenamefont {Julian},\ and\ \citenamefont
  {Lonzarich}}]{Pfleiderer01}%
  \BibitemOpen
  \bibfield  {author} {\bibinfo {author} {\bibfnamefont {C.}~\bibnamefont
  {Pfleiderer}}, \bibinfo {author} {\bibfnamefont {S.~R.}\ \bibnamefont
  {Julian}},\ and\ \bibinfo {author} {\bibfnamefont {G.~G.}\ \bibnamefont
  {Lonzarich}},\ }\bibfield  {title} {\bibinfo {title} {Non-\uppercase{F}ermi
  liquid nature of the normal state of itinerant-electron ferromagnets},\
  }\href {https://doi.org/10.1038/35106527} {\bibfield  {journal} {\bibinfo
  {journal} {Nature}\ }\textbf {\bibinfo {volume} {414}},\ \bibinfo {pages}
  {427} (\bibinfo {year} {2001})}\BibitemShut {NoStop}%
\bibitem [{\citenamefont {Doiron-Leyraud}\ \emph {et~al.}(2003)\citenamefont
  {Doiron-Leyraud}, \citenamefont {Walker}, \citenamefont {Taillefer},
  \citenamefont {Steiner}, \citenamefont {Julian},\ and\ \citenamefont
  {Lonzarich}}]{Doiron03}%
  \BibitemOpen
  \bibfield  {author} {\bibinfo {author} {\bibfnamefont {N.}~\bibnamefont
  {Doiron-Leyraud}}, \bibinfo {author} {\bibfnamefont {I.~R.}\ \bibnamefont
  {Walker}}, \bibinfo {author} {\bibfnamefont {L.}~\bibnamefont {Taillefer}},
  \bibinfo {author} {\bibfnamefont {M.~J.}\ \bibnamefont {Steiner}}, \bibinfo
  {author} {\bibfnamefont {S.~R.}\ \bibnamefont {Julian}},\ and\ \bibinfo
  {author} {\bibfnamefont {G.~G.}\ \bibnamefont {Lonzarich}},\ }\bibfield
  {title} {\bibinfo {title} {Fermi-liquid breakdown in the paramagnetic phase
  of a pure metal},\ }\href {https://doi.org/10.1038/nature01968} {\bibfield
  {journal} {\bibinfo  {journal} {Nature (London)}\ }\textbf {\bibinfo {volume}
  {425}},\ \bibinfo {pages} {595} (\bibinfo {year} {2003})}\BibitemShut
  {NoStop}%
\bibitem [{\citenamefont {Wilde}\ \emph {et~al.}(2021)\citenamefont {Wilde},
  \citenamefont {Dodenh\"oft}, \citenamefont {Niedermayr}, \citenamefont
  {Bauer}, \citenamefont {Hirschmann}, \citenamefont {Alpin}, \citenamefont
  {Schnyder},\ and\ \citenamefont {Pfleiderer}}]{Wilde21}%
  \BibitemOpen
  \bibfield  {author} {\bibinfo {author} {\bibfnamefont {M.~A.}\ \bibnamefont
  {Wilde}}, \bibinfo {author} {\bibfnamefont {M.}~\bibnamefont {Dodenh\"oft}},
  \bibinfo {author} {\bibfnamefont {A.}~\bibnamefont {Niedermayr}}, \bibinfo
  {author} {\bibfnamefont {A.}~\bibnamefont {Bauer}}, \bibinfo {author}
  {\bibfnamefont {M.~M.}\ \bibnamefont {Hirschmann}}, \bibinfo {author}
  {\bibfnamefont {K.}~\bibnamefont {Alpin}}, \bibinfo {author} {\bibfnamefont
  {A.~P.}\ \bibnamefont {Schnyder}},\ and\ \bibinfo {author} {\bibfnamefont
  {C.}~\bibnamefont {Pfleiderer}},\ }\bibfield  {title} {\bibinfo {title}
  {Symmetry-enforced topological nodal planes at the {F}ermi surface of a
  chiral magnet},\ }\href {https://doi.org/10.1038/s41586-021-03543-x}
  {\bibfield  {journal} {\bibinfo  {journal} {Nature}\ }\textbf {\bibinfo
  {volume} {594}},\ \bibinfo {pages} {374} (\bibinfo {year}
  {2021})}\BibitemShut {NoStop}%
\bibitem [{\citenamefont {Gendron}\ \emph {et~al.}(2022)\citenamefont
  {Gendron}, \citenamefont {Cliche},\ and\ \citenamefont {Amadon}}]{Gendron22}%
  \BibitemOpen
  \bibfield  {author} {\bibinfo {author} {\bibfnamefont {F.}~\bibnamefont
  {Gendron}}, \bibinfo {author} {\bibfnamefont {N.}~\bibnamefont {Cliche}},\
  and\ \bibinfo {author} {\bibfnamefont {B.}~\bibnamefont {Amadon}},\
  }\bibfield  {title} {\bibinfo {title} {Role of pressure on electronic,
  magnetic and structural properties at iron's {C}urie temperature: a {DFT} +
  {DMFT} study},\ }\href {https://doi.org/10.1088/1361-648X/ac8fd0} {\bibfield
  {journal} {\bibinfo  {journal} {J. Phys.: Condens. Matter}\ }\textbf
  {\bibinfo {volume} {34}},\ \bibinfo {pages} {464003} (\bibinfo {year}
  {2022})}\BibitemShut {NoStop}%
\bibitem [{\citenamefont {Cao}\ \emph {et~al.}(2024)\citenamefont {Cao},
  \citenamefont {Xu},\ and\ \citenamefont {Yang}}]{Cao24}%
  \BibitemOpen
  \bibfield  {author} {\bibinfo {author} {\bibfnamefont {Y.}~\bibnamefont
  {Cao}}, \bibinfo {author} {\bibfnamefont {Y.}~\bibnamefont {Xu}},\ and\
  \bibinfo {author} {\bibfnamefont {Y.-f.}\ \bibnamefont {Yang}},\ }\href
  {https://arxiv.org/abs/2405.01005} {\bibinfo {title} {Hundness and band
  renormalization in the kagome antiferromagnets {M}n$_3{X}$}} (\bibinfo {year}
  {2024}),\ \Eprint {https://arxiv.org/abs/2405.01005} {arXiv:2405.01005}
  \BibitemShut {NoStop}%
\bibitem [{\citenamefont {Takigawa}\ \emph {et~al.}(1980)\citenamefont
  {Takigawa}, \citenamefont {Yasuoka}, \citenamefont {Uemura}, \citenamefont
  {Hayano}, \citenamefont {Yamazaki},\ and\ \citenamefont
  {Ishikawa}}]{Takigawa80}%
  \BibitemOpen
  \bibfield  {author} {\bibinfo {author} {\bibfnamefont {M.}~\bibnamefont
  {Takigawa}}, \bibinfo {author} {\bibfnamefont {H.}~\bibnamefont {Yasuoka}},
  \bibinfo {author} {\bibfnamefont {Y.~J.}\ \bibnamefont {Uemura}}, \bibinfo
  {author} {\bibfnamefont {R.~S.}\ \bibnamefont {Hayano}}, \bibinfo {author}
  {\bibfnamefont {T.}~\bibnamefont {Yamazaki}},\ and\ \bibinfo {author}
  {\bibfnamefont {Y.}~\bibnamefont {Ishikawa}},\ }\bibfield  {title} {\bibinfo
  {title} {Positive muon spin rotation and relaxation studies in the helically
  ordered state of \uppercase{M}n\uppercase{S}i},\ }\href
  {https://doi.org/10.1143/JPSJ.49.1760} {\bibfield  {journal} {\bibinfo
  {journal} {J. Phys. Soc. Jpn.}\ }\textbf {\bibinfo {volume} {49}},\ \bibinfo
  {pages} {1760} (\bibinfo {year} {1980})}\BibitemShut {NoStop}%
\bibitem [{\citenamefont {Grigoriev}\ \emph {et~al.}(2006)\citenamefont
  {Grigoriev}, \citenamefont {Maleyev}, \citenamefont {Okorokov}, \citenamefont
  {Chetverikov}, \citenamefont {B\"oni}, \citenamefont {Georgii}, \citenamefont
  {Lamago}, \citenamefont {Eckerlebe},\ and\ \citenamefont
  {Pranzas}}]{Grigoriev06}%
  \BibitemOpen
  \bibfield  {author} {\bibinfo {author} {\bibfnamefont {S.~V.}\ \bibnamefont
  {Grigoriev}}, \bibinfo {author} {\bibfnamefont {S.~V.}\ \bibnamefont
  {Maleyev}}, \bibinfo {author} {\bibfnamefont {A.~I.}\ \bibnamefont
  {Okorokov}}, \bibinfo {author} {\bibfnamefont {Y.~O.}\ \bibnamefont
  {Chetverikov}}, \bibinfo {author} {\bibfnamefont {P.}~\bibnamefont {B\"oni}},
  \bibinfo {author} {\bibfnamefont {R.}~\bibnamefont {Georgii}}, \bibinfo
  {author} {\bibfnamefont {D.}~\bibnamefont {Lamago}}, \bibinfo {author}
  {\bibfnamefont {H.}~\bibnamefont {Eckerlebe}},\ and\ \bibinfo {author}
  {\bibfnamefont {K.}~\bibnamefont {Pranzas}},\ }\bibfield  {title} {\bibinfo
  {title} {Magnetic structure of \uppercase{M}n\uppercase{S}i under an applied
  field probed by polarized small-angle neutron scattering},\ }\href
  {https://doi.org/10.1103/PhysRevB.74.214414} {\bibfield  {journal} {\bibinfo
  {journal} {Phys. Rev. B}\ }\textbf {\bibinfo {volume} {74}},\ \bibinfo
  {pages} {214414} (\bibinfo {year} {2006})}\BibitemShut {NoStop}%
\bibitem [{\citenamefont {Kadono}\ \emph {et~al.}(1990)\citenamefont {Kadono},
  \citenamefont {Matsuzaki}, \citenamefont {Yamazaki}, \citenamefont
  {Kreitzman},\ and\ \citenamefont {Brewer}}]{Kadono90}%
  \BibitemOpen
  \bibfield  {author} {\bibinfo {author} {\bibfnamefont {R.}~\bibnamefont
  {Kadono}}, \bibinfo {author} {\bibfnamefont {T.}~\bibnamefont {Matsuzaki}},
  \bibinfo {author} {\bibfnamefont {T.}~\bibnamefont {Yamazaki}}, \bibinfo
  {author} {\bibfnamefont {S.~R.}\ \bibnamefont {Kreitzman}},\ and\ \bibinfo
  {author} {\bibfnamefont {J.~H.}\ \bibnamefont {Brewer}},\ }\bibfield  {title}
  {\bibinfo {title} {Spin dynamics of the itinerant helimagnet
  \uppercase{M}n\uppercase{S}i studied by positive muon spin relaxation},\
  }\href {https://doi.org/10.1103/PhysRevB.42.6515} {\bibfield  {journal}
  {\bibinfo  {journal} {Phys. Rev. B}\ }\textbf {\bibinfo {volume} {42}},\
  \bibinfo {pages} {6515} (\bibinfo {year} {1990})}\BibitemShut {NoStop}%
\bibitem [{\citenamefont {Bak}\ and\ \citenamefont {Jensen}(1980)}]{Bak80}%
  \BibitemOpen
  \bibfield  {author} {\bibinfo {author} {\bibfnamefont {P.}~\bibnamefont
  {Bak}}\ and\ \bibinfo {author} {\bibfnamefont {M.~H.}\ \bibnamefont
  {Jensen}},\ }\bibfield  {title} {\bibinfo {title} {Theory of helical magnetic
  structures and phase transitions in \uppercase{M}n\uppercase{S}i and
  \uppercase{F}e\uppercase{G}e},\ }\href
  {https://doi.org/10.1088/0022-3719/13/31/002} {\bibfield  {journal} {\bibinfo
   {journal} {J. Phys. C}\ }\textbf {\bibinfo {volume} {13}},\ \bibinfo {pages}
  {L881} (\bibinfo {year} {1980})}\BibitemShut {NoStop}%
\bibitem [{\citenamefont {Chaikin}\ and\ \citenamefont
  {Lubensky}(1995)}]{Chaikin95}%
  \BibitemOpen
  \bibfield  {author} {\bibinfo {author} {\bibfnamefont {P.~M.}\ \bibnamefont
  {Chaikin}}\ and\ \bibinfo {author} {\bibfnamefont {T.~C.}\ \bibnamefont
  {Lubensky}},\ }\href {https://doi.org/10.1017/CBO9780511813467} {\emph
  {\bibinfo {title} {Principles of Condensed Matter Physics}}}\ (\bibinfo
  {publisher} {Cambridge University Press},\ \bibinfo {address} {Cambridge},\
  \bibinfo {year} {1995})\BibitemShut {NoStop}%
\bibitem [{\citenamefont {Garst}\ \emph {et~al.}(2017)\citenamefont {Garst},
  \citenamefont {Waizner},\ and\ \citenamefont {Grundler}}]{Garst17}%
  \BibitemOpen
  \bibfield  {author} {\bibinfo {author} {\bibfnamefont {M.}~\bibnamefont
  {Garst}}, \bibinfo {author} {\bibfnamefont {J.}~\bibnamefont {Waizner}},\
  and\ \bibinfo {author} {\bibfnamefont {D.}~\bibnamefont {Grundler}},\
  }\bibfield  {title} {\bibinfo {title} {Collective spin excitations of helices
  and magnetic skyrmions: Review and perspectives of magnonics in
  non-centrosymmetric magnets},\ }\href
  {https://doi.org/10.1088/1361-6463/aa7573} {\bibfield  {journal} {\bibinfo
  {journal} {J. Phys. D}\ }\textbf {\bibinfo {volume} {50}},\ \bibinfo {pages}
  {293002} (\bibinfo {year} {2017})}\BibitemShut {NoStop}%
\bibitem [{\citenamefont {K\"ormann}\ \emph {et~al.}(2009)\citenamefont
  {K\"ormann}, \citenamefont {Dick}, \citenamefont {Hickel},\ and\
  \citenamefont {Neugebauer}}]{Kormann09}%
  \BibitemOpen
  \bibfield  {author} {\bibinfo {author} {\bibfnamefont {F.}~\bibnamefont
  {K\"ormann}}, \bibinfo {author} {\bibfnamefont {A.}~\bibnamefont {Dick}},
  \bibinfo {author} {\bibfnamefont {T.}~\bibnamefont {Hickel}},\ and\ \bibinfo
  {author} {\bibfnamefont {J.}~\bibnamefont {Neugebauer}},\ }\bibfield  {title}
  {\bibinfo {title} {Pressure dependence of the {C}urie temperature in bcc iron
  studied by {\sl ab initio\/} simulations},\ }\href
  {https://doi.org/10.1103/PhysRevB.79.184406} {\bibfield  {journal} {\bibinfo
  {journal} {Phys. Rev. B}\ }\textbf {\bibinfo {volume} {79}},\ \bibinfo
  {pages} {184406} (\bibinfo {year} {2009})}\BibitemShut {NoStop}%
\bibitem [{\citenamefont {Ishikawa}\ \emph {et~al.}(1977)\citenamefont
  {Ishikawa}, \citenamefont {Shirane}, \citenamefont {Tarvin},\ and\
  \citenamefont {Kohgi}}]{Ishikawa77}%
  \BibitemOpen
  \bibfield  {author} {\bibinfo {author} {\bibfnamefont {Y.}~\bibnamefont
  {Ishikawa}}, \bibinfo {author} {\bibfnamefont {G.}~\bibnamefont {Shirane}},
  \bibinfo {author} {\bibfnamefont {J.~A.}\ \bibnamefont {Tarvin}},\ and\
  \bibinfo {author} {\bibfnamefont {M.}~\bibnamefont {Kohgi}},\ }\bibfield
  {title} {\bibinfo {title} {Magnetic excitations in the weak itinerant
  ferromagnet \uppercase{M}n\uppercase{S}i},\ }\href
  {https://doi.org/10.1103/PhysRevB.16.4956} {\bibfield  {journal} {\bibinfo
  {journal} {Phys. Rev. B}\ }\textbf {\bibinfo {volume} {16}},\ \bibinfo
  {pages} {4956} (\bibinfo {year} {1977})}\BibitemShut {NoStop}%
\bibitem [{\citenamefont {Jin}\ \emph {et~al.}(2023)\citenamefont {Jin},
  \citenamefont {Li}, \citenamefont {Hu}, \citenamefont {Hu}, \citenamefont
  {Liu}, \citenamefont {Iida}, \citenamefont {Kamazawa}, \citenamefont {Stone},
  \citenamefont {Kolesnikov}, \citenamefont {Abernathy}, \citenamefont {Zhang},
  \citenamefont {Chen}, \citenamefont {Wang}, \citenamefont {Fang},
  \citenamefont {Wu}, \citenamefont {Zaliznyak}, \citenamefont {Tranquada},\
  and\ \citenamefont {Li}}]{Jin23}%
  \BibitemOpen
  \bibfield  {author} {\bibinfo {author} {\bibfnamefont {Z.}~\bibnamefont
  {Jin}}, \bibinfo {author} {\bibfnamefont {Y.}~\bibnamefont {Li}}, \bibinfo
  {author} {\bibfnamefont {Z.}~\bibnamefont {Hu}}, \bibinfo {author}
  {\bibfnamefont {B.}~\bibnamefont {Hu}}, \bibinfo {author} {\bibfnamefont
  {Y.}~\bibnamefont {Liu}}, \bibinfo {author} {\bibfnamefont {K.}~\bibnamefont
  {Iida}}, \bibinfo {author} {\bibfnamefont {K.}~\bibnamefont {Kamazawa}},
  \bibinfo {author} {\bibfnamefont {M.~B.}\ \bibnamefont {Stone}}, \bibinfo
  {author} {\bibfnamefont {A.~I.}\ \bibnamefont {Kolesnikov}}, \bibinfo
  {author} {\bibfnamefont {D.~L.}\ \bibnamefont {Abernathy}}, \bibinfo {author}
  {\bibfnamefont {X.}~\bibnamefont {Zhang}}, \bibinfo {author} {\bibfnamefont
  {H.}~\bibnamefont {Chen}}, \bibinfo {author} {\bibfnamefont {Y.}~\bibnamefont
  {Wang}}, \bibinfo {author} {\bibfnamefont {C.}~\bibnamefont {Fang}}, \bibinfo
  {author} {\bibfnamefont {B.}~\bibnamefont {Wu}}, \bibinfo {author}
  {\bibfnamefont {I.~A.}\ \bibnamefont {Zaliznyak}}, \bibinfo {author}
  {\bibfnamefont {J.~M.}\ \bibnamefont {Tranquada}},\ and\ \bibinfo {author}
  {\bibfnamefont {Y.}~\bibnamefont {Li}},\ }\bibfield  {title} {\bibinfo
  {title} {Magnetic molecular orbitals in {M}n{S}i},\ }\href
  {https://doi.org/10.1126/sciadv.add5239} {\bibfield  {journal} {\bibinfo
  {journal} {Sci. Adv.}\ }\textbf {\bibinfo {volume} {9}},\ \bibinfo {pages}
  {eadd5239} (\bibinfo {year} {2023})}\BibitemShut {NoStop}%
\end{thebibliography}%


\begin{thebibliography}{26}%
\makeatletter
\providecommand \@ifxundefined [1]{%
 \@ifx{#1\undefined}
}%
\providecommand \@ifnum [1]{%
 \ifnum #1\expandafter \@firstoftwo
 \else \expandafter \@secondoftwo
 \fi
}%
\providecommand \@ifx [1]{%
 \ifx #1\expandafter \@firstoftwo
 \else \expandafter \@secondoftwo
 \fi
}%
\providecommand \natexlab [1]{#1}%
\providecommand \enquote  [1]{``#1''}%
\providecommand \bibnamefont  [1]{#1}%
\providecommand \bibfnamefont [1]{#1}%
\providecommand \citenamefont [1]{#1}%
\providecommand \href@noop [0]{\@secondoftwo}%
\providecommand \href [0]{\begingroup \@sanitize@url \@href}%
\providecommand \@href[1]{\@@startlink{#1}\@@href}%
\providecommand \@@href[1]{\endgroup#1\@@endlink}%
\providecommand \@sanitize@url [0]{\catcode `\\12\catcode `\$12\catcode
  `\&12\catcode `\#12\catcode `\^12\catcode `\_12\catcode `\%12\relax}%
\providecommand \@@startlink[1]{}%
\providecommand \@@endlink[0]{}%
\providecommand \url  [0]{\begingroup\@sanitize@url \@url }%
\providecommand \@url [1]{\endgroup\@href {#1}{\urlprefix }}%
\providecommand \urlprefix  [0]{URL }%
\providecommand \Eprint [0]{\href }%
\providecommand \doibase [0]{https://doi.org/}%
\providecommand \selectlanguage [0]{\@gobble}%
\providecommand \bibinfo  [0]{\@secondoftwo}%
\providecommand \bibfield  [0]{\@secondoftwo}%
\providecommand \translation [1]{[#1]}%
\providecommand \BibitemOpen [0]{}%
\providecommand \bibitemStop [0]{}%
\providecommand \bibitemNoStop [0]{.\EOS\space}%
\providecommand \EOS [0]{\spacefactor3000\relax}%
\providecommand \BibitemShut  [1]{\csname bibitem#1\endcsname}%
\let\auto@bib@innerbib\@empty
\bibitem [{\citenamefont {Bor\'en}(1933)}]{Boren33}%
  \BibitemOpen
  \bibfield  {author} {\bibinfo {author} {\bibfnamefont {B.}~\bibnamefont
  {Bor\'en}},\ }\bibfield  {title} {\bibinfo {title} {Roentgenuntersuchung der
  legierungen von silicium mit chrom, mangan, kobalt und nickel},\ }\href@noop
  {} {\bibfield  {journal} {\bibinfo  {journal} {Ark. Kemi Mineral. Geol.}\
  }\textbf {\bibinfo {volume} {11A}},\ \bibinfo {pages} {1} (\bibinfo {year}
  {1933})}\BibitemShut {NoStop}%
\bibitem [{\citenamefont {Yaouanc}\ and\ \citenamefont {{Dalmas de
  R\'eotier}}(2011)}]{Yaouanc11}%
  \BibitemOpen
  \bibfield  {author} {\bibinfo {author} {\bibfnamefont {A.}~\bibnamefont
  {Yaouanc}}\ and\ \bibinfo {author} {\bibfnamefont {P.}~\bibnamefont {{Dalmas
  de R\'eotier}}},\ }\href@noop {} {\emph {\bibinfo {title} {Muon Spin
  Rotation, Relaxation, and Resonance: Applications to Condensed Matter}}}\
  (\bibinfo  {publisher} {Oxford University Press},\ \bibinfo {address}
  {Oxford},\ \bibinfo {year} {2011})\BibitemShut {NoStop}%
\bibitem [{\citenamefont {Amato}\ and\ \citenamefont
  {Morenzoni}(2024)}]{Amato24}%
  \BibitemOpen
  \bibfield  {author} {\bibinfo {author} {\bibfnamefont {A.}~\bibnamefont
  {Amato}}\ and\ \bibinfo {author} {\bibfnamefont {E.}~\bibnamefont
  {Morenzoni}},\ }\href@noop {} {\emph {\bibinfo {title} {Introduction to muon
  spin spectroscopy}}},\ Lecture notes in physics\ (\bibinfo  {publisher}
  {Springer},\ \bibinfo {year} {2024})\BibitemShut {NoStop}%
\bibitem [{\citenamefont {Takigawa}\ \emph {et~al.}(1980)\citenamefont
  {Takigawa}, \citenamefont {Yasuoka}, \citenamefont {Uemura}, \citenamefont
  {Hayano}, \citenamefont {Yamazaki},\ and\ \citenamefont
  {Ishikawa}}]{Takigawa80}%
  \BibitemOpen
  \bibfield  {author} {\bibinfo {author} {\bibfnamefont {M.}~\bibnamefont
  {Takigawa}}, \bibinfo {author} {\bibfnamefont {H.}~\bibnamefont {Yasuoka}},
  \bibinfo {author} {\bibfnamefont {Y.~J.}\ \bibnamefont {Uemura}}, \bibinfo
  {author} {\bibfnamefont {R.~S.}\ \bibnamefont {Hayano}}, \bibinfo {author}
  {\bibfnamefont {T.}~\bibnamefont {Yamazaki}},\ and\ \bibinfo {author}
  {\bibfnamefont {Y.}~\bibnamefont {Ishikawa}},\ }\bibfield  {title} {\bibinfo
  {title} {Positive muon spin rotation and relaxation studies in the helically
  ordered state of \uppercase{M}n\uppercase{S}i},\ }\href
  {https://doi.org/10.1143/JPSJ.49.1760} {\bibfield  {journal} {\bibinfo
  {journal} {J. Phys. Soc. Jpn.}\ }\textbf {\bibinfo {volume} {49}},\ \bibinfo
  {pages} {1760} (\bibinfo {year} {1980})}\BibitemShut {NoStop}%
\bibitem [{\citenamefont {Dalmas~de R\'eotier}\ \emph
  {et~al.}(2016)\citenamefont {Dalmas~de R\'eotier}, \citenamefont
  {Maisuradze}, \citenamefont {Yaouanc}, \citenamefont {Roessli}, \citenamefont
  {Amato}, \citenamefont {Andreica},\ and\ \citenamefont
  {Lapertot}}]{Dalmas16}%
  \BibitemOpen
  \bibfield  {author} {\bibinfo {author} {\bibfnamefont {P.}~\bibnamefont
  {Dalmas~de R\'eotier}}, \bibinfo {author} {\bibfnamefont {A.}~\bibnamefont
  {Maisuradze}}, \bibinfo {author} {\bibfnamefont {A.}~\bibnamefont {Yaouanc}},
  \bibinfo {author} {\bibfnamefont {B.}~\bibnamefont {Roessli}}, \bibinfo
  {author} {\bibfnamefont {A.}~\bibnamefont {Amato}}, \bibinfo {author}
  {\bibfnamefont {D.}~\bibnamefont {Andreica}},\ and\ \bibinfo {author}
  {\bibfnamefont {G.}~\bibnamefont {Lapertot}},\ }\bibfield  {title} {\bibinfo
  {title} {Determination of the zero-field magnetic structure of the helimagnet
  \uppercase{M}n\uppercase{S}i at low temperature},\ }\href
  {https://doi.org/10.1103/PhysRevB.93.144419} {\bibfield  {journal} {\bibinfo
  {journal} {Phys. Rev. B}\ }\textbf {\bibinfo {volume} {93}},\ \bibinfo
  {pages} {144419} (\bibinfo {year} {2016})}\BibitemShut {NoStop}%
\bibitem [{\citenamefont {Andreica}\ \emph {et~al.}(2010)\citenamefont
  {Andreica}, \citenamefont {{Dalmas de R\'eotier}}, \citenamefont {Yaouanc},
  \citenamefont {Amato},\ and\ \citenamefont {Lapertot}}]{Andreica10}%
  \BibitemOpen
  \bibfield  {author} {\bibinfo {author} {\bibfnamefont {D.}~\bibnamefont
  {Andreica}}, \bibinfo {author} {\bibfnamefont {P.}~\bibnamefont {{Dalmas de
  R\'eotier}}}, \bibinfo {author} {\bibfnamefont {A.}~\bibnamefont {Yaouanc}},
  \bibinfo {author} {\bibfnamefont {A.}~\bibnamefont {Amato}},\ and\ \bibinfo
  {author} {\bibfnamefont {G.}~\bibnamefont {Lapertot}},\ }\bibfield  {title}
  {\bibinfo {title} {Absence of magnetic phase separation in
  \uppercase{M}n\uppercase{S}i under pressure},\ }\href
  {https://doi.org/10.1103/PhysRevB.81.060412} {\bibfield  {journal} {\bibinfo
  {journal} {Phys. Rev. B}\ }\textbf {\bibinfo {volume} {81}},\ \bibinfo
  {pages} {060412(R)} (\bibinfo {year} {2010})}\BibitemShut {NoStop}%
\bibitem [{\citenamefont {Pfleiderer}\ \emph {et~al.}(1997)\citenamefont
  {Pfleiderer}, \citenamefont {McMullan}, \citenamefont {Julian},\ and\
  \citenamefont {Lonzarich}}]{Pfleiderer97}%
  \BibitemOpen
  \bibfield  {author} {\bibinfo {author} {\bibfnamefont {C.}~\bibnamefont
  {Pfleiderer}}, \bibinfo {author} {\bibfnamefont {G.~J.}\ \bibnamefont
  {McMullan}}, \bibinfo {author} {\bibfnamefont {S.~R.}\ \bibnamefont
  {Julian}},\ and\ \bibinfo {author} {\bibfnamefont {G.~G.}\ \bibnamefont
  {Lonzarich}},\ }\bibfield  {title} {\bibinfo {title} {Magnetic quantum phase
  transition in \uppercase{M}n\uppercase{S}i under hydrostatic pressure},\
  }\href {https://doi.org/10.1103/PhysRevB.55.8330} {\bibfield  {journal}
  {\bibinfo  {journal} {Phys. Rev. B}\ }\textbf {\bibinfo {volume} {55}},\
  \bibinfo {pages} {8330} (\bibinfo {year} {1997})}\BibitemShut {NoStop}%
\bibitem [{\citenamefont {F{\aa}k}\ \emph {et~al.}(2005)\citenamefont
  {F{\aa}k}, \citenamefont {Sadykov}, \citenamefont {Flouquet},\ and\
  \citenamefont {Lapertot}}]{Fak05}%
  \BibitemOpen
  \bibfield  {author} {\bibinfo {author} {\bibfnamefont {B.}~\bibnamefont
  {F{\aa}k}}, \bibinfo {author} {\bibfnamefont {R.~A.}\ \bibnamefont
  {Sadykov}}, \bibinfo {author} {\bibfnamefont {J.}~\bibnamefont {Flouquet}},\
  and\ \bibinfo {author} {\bibfnamefont {G.}~\bibnamefont {Lapertot}},\
  }\bibfield  {title} {\bibinfo {title} {Pressure dependence of the magnetic
  structure of the itinerant electron magnet \uppercase{M}n\uppercase{S}i},\
  }\href {https://doi.org/10.1088/0953-8984/17/10/018} {\bibfield  {journal}
  {\bibinfo  {journal} {J. Phys.: Condens. Matter}\ }\textbf {\bibinfo {volume}
  {17}},\ \bibinfo {pages} {1635} (\bibinfo {year} {2005})}\BibitemShut
  {NoStop}%
\bibitem [{\citenamefont {Amato}\ \emph {et~al.}(2014)\citenamefont {Amato},
  \citenamefont {{Dalmas de R\'eotier}}, \citenamefont {Andreica},
  \citenamefont {Yaouanc}, \citenamefont {Suter}, \citenamefont {Lapertot},
  \citenamefont {Pop}, \citenamefont {Morenzoni}, \citenamefont {Bonf\`a},
  \citenamefont {Bernardini},\ and\ \citenamefont {De~Renzi}}]{Amato14}%
  \BibitemOpen
  \bibfield  {author} {\bibinfo {author} {\bibfnamefont {A.}~\bibnamefont
  {Amato}}, \bibinfo {author} {\bibfnamefont {P.}~\bibnamefont {{Dalmas de
  R\'eotier}}}, \bibinfo {author} {\bibfnamefont {D.}~\bibnamefont {Andreica}},
  \bibinfo {author} {\bibfnamefont {A.}~\bibnamefont {Yaouanc}}, \bibinfo
  {author} {\bibfnamefont {A.}~\bibnamefont {Suter}}, \bibinfo {author}
  {\bibfnamefont {G.}~\bibnamefont {Lapertot}}, \bibinfo {author}
  {\bibfnamefont {I.~M.}\ \bibnamefont {Pop}}, \bibinfo {author} {\bibfnamefont
  {E.}~\bibnamefont {Morenzoni}}, \bibinfo {author} {\bibfnamefont
  {P.}~\bibnamefont {Bonf\`a}}, \bibinfo {author} {\bibfnamefont
  {F.}~\bibnamefont {Bernardini}},\ and\ \bibinfo {author} {\bibfnamefont
  {R.}~\bibnamefont {De~Renzi}},\ }\bibfield  {title} {\bibinfo {title}
  {Understanding the $\mu$\uppercase{SR} spectra of
  \uppercase{M}n\uppercase{S}i without magnetic polarons},\ }\href
  {https://doi.org/10.1103/PhysRevB.89.184425} {\bibfield  {journal} {\bibinfo
  {journal} {Phys. Rev. B}\ }\textbf {\bibinfo {volume} {89}},\ \bibinfo
  {pages} {184425} (\bibinfo {year} {2014})}\BibitemShut {NoStop}%
\bibitem [{\citenamefont {{Dalmas de R\'eotier}}\ \emph
  {et~al.}(2018)\citenamefont {{Dalmas de R\'eotier}}, \citenamefont {Yaouanc},
  \citenamefont {Amato}, \citenamefont {Maisuradze}, \citenamefont {Andreica},
  \citenamefont {Roessli}, \citenamefont {Goko}, \citenamefont {Scheuermann},\
  and\ \citenamefont {Lapertot}}]{Dalmas18}%
  \BibitemOpen
  \bibfield  {author} {\bibinfo {author} {\bibfnamefont {P.}~\bibnamefont
  {{Dalmas de R\'eotier}}}, \bibinfo {author} {\bibfnamefont {A.}~\bibnamefont
  {Yaouanc}}, \bibinfo {author} {\bibfnamefont {A.}~\bibnamefont {Amato}},
  \bibinfo {author} {\bibfnamefont {A.}~\bibnamefont {Maisuradze}}, \bibinfo
  {author} {\bibfnamefont {D.}~\bibnamefont {Andreica}}, \bibinfo {author}
  {\bibfnamefont {B.}~\bibnamefont {Roessli}}, \bibinfo {author} {\bibfnamefont
  {T.}~\bibnamefont {Goko}}, \bibinfo {author} {\bibfnamefont {R.}~\bibnamefont
  {Scheuermann}},\ and\ \bibinfo {author} {\bibfnamefont {G.}~\bibnamefont
  {Lapertot}},\ }\bibfield  {title} {\bibinfo {title} {On the robustness of the
  \uppercase{M}n\uppercase{S}i magnetic structure determined by muon spin
  rotation},\ }\href {https://doi.org/10.3390/qubs2030019} {\bibfield
  {journal} {\bibinfo  {journal} {Quantum Beam Sci.}\ }\textbf {\bibinfo
  {volume} {2}},\ \bibinfo {pages} {19} (\bibinfo {year} {2018})}\BibitemShut
  {NoStop}%
\bibitem [{\citenamefont {Grigoriev}\ \emph {et~al.}(2006)\citenamefont
  {Grigoriev}, \citenamefont {Maleyev}, \citenamefont {Okorokov}, \citenamefont
  {Chetverikov}, \citenamefont {B\"oni}, \citenamefont {Georgii}, \citenamefont
  {Lamago}, \citenamefont {Eckerlebe},\ and\ \citenamefont
  {Pranzas}}]{Grigoriev06}%
  \BibitemOpen
  \bibfield  {author} {\bibinfo {author} {\bibfnamefont {S.~V.}\ \bibnamefont
  {Grigoriev}}, \bibinfo {author} {\bibfnamefont {S.~V.}\ \bibnamefont
  {Maleyev}}, \bibinfo {author} {\bibfnamefont {A.~I.}\ \bibnamefont
  {Okorokov}}, \bibinfo {author} {\bibfnamefont {Y.~O.}\ \bibnamefont
  {Chetverikov}}, \bibinfo {author} {\bibfnamefont {P.}~\bibnamefont {B\"oni}},
  \bibinfo {author} {\bibfnamefont {R.}~\bibnamefont {Georgii}}, \bibinfo
  {author} {\bibfnamefont {D.}~\bibnamefont {Lamago}}, \bibinfo {author}
  {\bibfnamefont {H.}~\bibnamefont {Eckerlebe}},\ and\ \bibinfo {author}
  {\bibfnamefont {K.}~\bibnamefont {Pranzas}},\ }\bibfield  {title} {\bibinfo
  {title} {Magnetic structure of \uppercase{M}n\uppercase{S}i under an applied
  field probed by polarized small-angle neutron scattering},\ }\href
  {https://doi.org/10.1103/PhysRevB.74.214414} {\bibfield  {journal} {\bibinfo
  {journal} {Phys. Rev. B}\ }\textbf {\bibinfo {volume} {74}},\ \bibinfo
  {pages} {214414} (\bibinfo {year} {2006})}\BibitemShut {NoStop}%
\bibitem [{\citenamefont {Chizhikov}\ and\ \citenamefont
  {Dmitrienko}(2012)}]{Chizhikov12}%
  \BibitemOpen
  \bibfield  {author} {\bibinfo {author} {\bibfnamefont {V.~A.}\ \bibnamefont
  {Chizhikov}}\ and\ \bibinfo {author} {\bibfnamefont {V.~E.}\ \bibnamefont
  {Dmitrienko}},\ }\bibfield  {title} {\bibinfo {title} {Frustrated magnetic
  helices in \uppercase{M}n\uppercase{S}i-type crystals},\ }\href
  {https://doi.org/10.1103/PhysRevB.85.014421} {\bibfield  {journal} {\bibinfo
  {journal} {Phys. Rev. B}\ }\textbf {\bibinfo {volume} {85}},\ \bibinfo
  {pages} {014421} (\bibinfo {year} {2012})}\BibitemShut {NoStop}%
\bibitem [{\citenamefont {Dalmas~de R\'eotier}\ \emph
  {et~al.}(2024)\citenamefont {Dalmas~de R\'eotier}, \citenamefont {Yaouanc},
  \citenamefont {Lapertot}, \citenamefont {Wang}, \citenamefont {Amato},\ and\
  \citenamefont {Andreica}}]{Dalmas24}%
  \BibitemOpen
  \bibfield  {author} {\bibinfo {author} {\bibfnamefont {P.}~\bibnamefont
  {Dalmas~de R\'eotier}}, \bibinfo {author} {\bibfnamefont {A.}~\bibnamefont
  {Yaouanc}}, \bibinfo {author} {\bibfnamefont {G.}~\bibnamefont {Lapertot}},
  \bibinfo {author} {\bibfnamefont {C.}~\bibnamefont {Wang}}, \bibinfo {author}
  {\bibfnamefont {A.}~\bibnamefont {Amato}},\ and\ \bibinfo {author}
  {\bibfnamefont {D.}~\bibnamefont {Andreica}},\ }\bibfield  {title} {\bibinfo
  {title} {Experimental determination of the spin {H}amiltonian of the cubic
  chiral magnet {M}n{S}i},\ }\href
  {https://doi.org/10.1103/PhysRevB.109.L020408} {\bibfield  {journal}
  {\bibinfo  {journal} {Phys. Rev. B}\ }\textbf {\bibinfo {volume} {109}},\
  \bibinfo {pages} {L020408} (\bibinfo {year} {2024})}\BibitemShut {NoStop}%
\bibitem [{\citenamefont {Yaouanc}\ \emph {et~al.}(2020)\citenamefont
  {Yaouanc}, \citenamefont {Dalmas~de R\'eotier}, \citenamefont {Roessli},
  \citenamefont {Maisuradze}, \citenamefont {Amato}, \citenamefont {Andreica},\
  and\ \citenamefont {Lapertot}}]{Yaouanc20}%
  \BibitemOpen
  \bibfield  {author} {\bibinfo {author} {\bibfnamefont {A.}~\bibnamefont
  {Yaouanc}}, \bibinfo {author} {\bibfnamefont {P.}~\bibnamefont {Dalmas~de
  R\'eotier}}, \bibinfo {author} {\bibfnamefont {B.}~\bibnamefont {Roessli}},
  \bibinfo {author} {\bibfnamefont {A.}~\bibnamefont {Maisuradze}}, \bibinfo
  {author} {\bibfnamefont {A.}~\bibnamefont {Amato}}, \bibinfo {author}
  {\bibfnamefont {D.}~\bibnamefont {Andreica}},\ and\ \bibinfo {author}
  {\bibfnamefont {G.}~\bibnamefont {Lapertot}},\ }\bibfield  {title} {\bibinfo
  {title} {Dual nature of magnetism in \uppercase{M}n\uppercase{S}i},\ }\href
  {https://doi.org/10.1103/PhysRevResearch.2.013029} {\bibfield  {journal}
  {\bibinfo  {journal} {Phys. Rev. Research}\ }\textbf {\bibinfo {volume}
  {2}},\ \bibinfo {pages} {013029} (\bibinfo {year} {2020})}\BibitemShut
  {NoStop}%
\bibitem [{\citenamefont {Bannenberg}\ \emph {et~al.}(2019)\citenamefont
  {Bannenberg}, \citenamefont {Sadykov}, \citenamefont {Dalgliesh},
  \citenamefont {Goodway}, \citenamefont {Schlagel}, \citenamefont {Lograsso},
  \citenamefont {Falus}, \citenamefont {Leli\`evre-Berna}, \citenamefont
  {Leonov},\ and\ \citenamefont {Pappas}}]{Bannenberg19}%
  \BibitemOpen
  \bibfield  {author} {\bibinfo {author} {\bibfnamefont {L.~J.}\ \bibnamefont
  {Bannenberg}}, \bibinfo {author} {\bibfnamefont {R.}~\bibnamefont {Sadykov}},
  \bibinfo {author} {\bibfnamefont {R.~M.}\ \bibnamefont {Dalgliesh}}, \bibinfo
  {author} {\bibfnamefont {C.}~\bibnamefont {Goodway}}, \bibinfo {author}
  {\bibfnamefont {D.~L.}\ \bibnamefont {Schlagel}}, \bibinfo {author}
  {\bibfnamefont {T.~A.}\ \bibnamefont {Lograsso}}, \bibinfo {author}
  {\bibfnamefont {P.}~\bibnamefont {Falus}}, \bibinfo {author} {\bibfnamefont
  {E.}~\bibnamefont {Leli\`evre-Berna}}, \bibinfo {author} {\bibfnamefont
  {A.~O.}\ \bibnamefont {Leonov}},\ and\ \bibinfo {author} {\bibfnamefont
  {C.}~\bibnamefont {Pappas}},\ }\bibfield  {title} {\bibinfo {title}
  {Skyrmions and spirals in \uppercase{M}n\uppercase{S}i under hydrostatic
  pressure},\ }\href {https://doi.org/10.1103/PhysRevB.100.054447} {\bibfield
  {journal} {\bibinfo  {journal} {Phys. Rev. B}\ }\textbf {\bibinfo {volume}
  {100}},\ \bibinfo {pages} {054447} (\bibinfo {year} {2019})}\BibitemShut
  {NoStop}%
\bibitem [{\citenamefont {Kadono}\ \emph {et~al.}(1990)\citenamefont {Kadono},
  \citenamefont {Matsuzaki}, \citenamefont {Yamazaki}, \citenamefont
  {Kreitzman},\ and\ \citenamefont {Brewer}}]{Kadono90}%
  \BibitemOpen
  \bibfield  {author} {\bibinfo {author} {\bibfnamefont {R.}~\bibnamefont
  {Kadono}}, \bibinfo {author} {\bibfnamefont {T.}~\bibnamefont {Matsuzaki}},
  \bibinfo {author} {\bibfnamefont {T.}~\bibnamefont {Yamazaki}}, \bibinfo
  {author} {\bibfnamefont {S.~R.}\ \bibnamefont {Kreitzman}},\ and\ \bibinfo
  {author} {\bibfnamefont {J.~H.}\ \bibnamefont {Brewer}},\ }\bibfield  {title}
  {\bibinfo {title} {Spin dynamics of the itinerant helimagnet
  \uppercase{M}n\uppercase{S}i studied by positive muon spin relaxation},\
  }\href {https://doi.org/10.1103/PhysRevB.42.6515} {\bibfield  {journal}
  {\bibinfo  {journal} {Phys. Rev. B}\ }\textbf {\bibinfo {volume} {42}},\
  \bibinfo {pages} {6515} (\bibinfo {year} {1990})}\BibitemShut {NoStop}%
\bibitem [{\citenamefont {Maleyev}(2006)}]{Maleyev06}%
  \BibitemOpen
  \bibfield  {author} {\bibinfo {author} {\bibfnamefont {S.~V.}\ \bibnamefont
  {Maleyev}},\ }\bibfield  {title} {\bibinfo {title} {Cubic magnets with
  \uppercase{D}zyaloshinskii-\uppercase{M}oriya interaction at low
  temperature},\ }\href {https://doi.org/10.1103/PhysRevB.73.174402} {\bibfield
   {journal} {\bibinfo  {journal} {Phys. Rev. B}\ }\textbf {\bibinfo {volume}
  {73}},\ \bibinfo {pages} {174402} (\bibinfo {year} {2006})}\BibitemShut
  {NoStop}%
\bibitem [{\citenamefont {Belitz}\ \emph {et~al.}(2006)\citenamefont {Belitz},
  \citenamefont {Kirkpatrick},\ and\ \citenamefont {Rosch}}]{Belitz06}%
  \BibitemOpen
  \bibfield  {author} {\bibinfo {author} {\bibfnamefont {D.}~\bibnamefont
  {Belitz}}, \bibinfo {author} {\bibfnamefont {T.~R.}\ \bibnamefont
  {Kirkpatrick}},\ and\ \bibinfo {author} {\bibfnamefont {A.}~\bibnamefont
  {Rosch}},\ }\bibfield  {title} {\bibinfo {title} {Theory of helimagnons in
  itinerant quantum systems},\ }\href
  {https://doi.org/10.1103/PhysRevB.73.054431} {\bibfield  {journal} {\bibinfo
  {journal} {Phys. Rev. B}\ }\textbf {\bibinfo {volume} {73}},\ \bibinfo
  {pages} {054431} (\bibinfo {year} {2006})}\BibitemShut {NoStop}%
\bibitem [{\citenamefont {Bak}\ and\ \citenamefont {Jensen}(1980)}]{Bak80}%
  \BibitemOpen
  \bibfield  {author} {\bibinfo {author} {\bibfnamefont {P.}~\bibnamefont
  {Bak}}\ and\ \bibinfo {author} {\bibfnamefont {M.~H.}\ \bibnamefont
  {Jensen}},\ }\bibfield  {title} {\bibinfo {title} {Theory of helical magnetic
  structures and phase transitions in \uppercase{M}n\uppercase{S}i and
  \uppercase{F}e\uppercase{G}e},\ }\href
  {https://doi.org/10.1088/0022-3719/13/31/002} {\bibfield  {journal} {\bibinfo
   {journal} {J. Phys. C}\ }\textbf {\bibinfo {volume} {13}},\ \bibinfo {pages}
  {L881} (\bibinfo {year} {1980})}\BibitemShut {NoStop}%
\bibitem [{\citenamefont {Chaikin}\ and\ \citenamefont
  {Lubensky}(1995)}]{Chaikin95}%
  \BibitemOpen
  \bibfield  {author} {\bibinfo {author} {\bibfnamefont {P.~M.}\ \bibnamefont
  {Chaikin}}\ and\ \bibinfo {author} {\bibfnamefont {T.~C.}\ \bibnamefont
  {Lubensky}},\ }\href {https://doi.org/10.1017/CBO9780511813467} {\emph
  {\bibinfo {title} {Principles of Condensed Matter Physics}}}\ (\bibinfo
  {publisher} {Cambridge University Press},\ \bibinfo {address} {Cambridge},\
  \bibinfo {year} {1995})\BibitemShut {NoStop}%
\bibitem [{\citenamefont {Garst}\ \emph {et~al.}(2017)\citenamefont {Garst},
  \citenamefont {Waizner},\ and\ \citenamefont {Grundler}}]{Garst17}%
  \BibitemOpen
  \bibfield  {author} {\bibinfo {author} {\bibfnamefont {M.}~\bibnamefont
  {Garst}}, \bibinfo {author} {\bibfnamefont {J.}~\bibnamefont {Waizner}},\
  and\ \bibinfo {author} {\bibfnamefont {D.}~\bibnamefont {Grundler}},\
  }\bibfield  {title} {\bibinfo {title} {Collective spin excitations of helices
  and magnetic skyrmions: review and perspectives of magnonics in
  non-centrosymmetric magnets},\ }\href
  {https://doi.org/10.1088/1361-6463/aa7573} {\bibfield  {journal} {\bibinfo
  {journal} {J. Phys. D: Applied Physics}\ }\textbf {\bibinfo {volume} {50}},\
  \bibinfo {pages} {293002} (\bibinfo {year} {2017})}\BibitemShut {NoStop}%
\bibitem [{\citenamefont {{Dalmas de R\'eotier}}\ and\ \citenamefont
  {Yaouanc}(2021)}]{Dalmas21}%
  \BibitemOpen
  \bibfield  {author} {\bibinfo {author} {\bibfnamefont {P.}~\bibnamefont
  {{Dalmas de R\'eotier}}}\ and\ \bibinfo {author} {\bibfnamefont
  {A.}~\bibnamefont {Yaouanc}},\ }\bibfield  {title} {\bibinfo {title}
  {Zero-field $^{29}$\uppercase{S}i nuclear magnetic resonance signature of
  helimagnons in \uppercase{M}n\uppercase{S}i},\ }\href
  {https://doi.org/10.1016/j.jmmm.2021.168086} {\bibfield  {journal} {\bibinfo
  {journal} {J. Magn. Magn. Mater.}\ }\textbf {\bibinfo {volume} {537}},\
  \bibinfo {pages} {168086} (\bibinfo {year} {2021})}\BibitemShut {NoStop}%
\bibitem [{\citenamefont {van Kranendonk}\ and\ \citenamefont {van
  Vleck}(1958)}]{vanKranendonk58}%
  \BibitemOpen
  \bibfield  {author} {\bibinfo {author} {\bibfnamefont {J.}~\bibnamefont {van
  Kranendonk}}\ and\ \bibinfo {author} {\bibfnamefont {J.~H.}\ \bibnamefont
  {van Vleck}},\ }\bibfield  {title} {\bibinfo {title} {Spin waves},\ }\href
  {https://doi.org/10.1103/RevModPhys.30.1} {\bibfield  {journal} {\bibinfo
  {journal} {Rev. Mod. Phys.}\ }\textbf {\bibinfo {volume} {30}},\ \bibinfo
  {pages} {1} (\bibinfo {year} {1958})}\BibitemShut {NoStop}%
\bibitem [{\citenamefont {Wang}\ \emph {et~al.}(1982)\citenamefont {Wang},
  \citenamefont {Prange},\ and\ \citenamefont {Korenman}}]{Wang82}%
  \BibitemOpen
  \bibfield  {author} {\bibinfo {author} {\bibfnamefont {C.~S.}\ \bibnamefont
  {Wang}}, \bibinfo {author} {\bibfnamefont {R.~E.}\ \bibnamefont {Prange}},\
  and\ \bibinfo {author} {\bibfnamefont {V.}~\bibnamefont {Korenman}},\
  }\bibfield  {title} {\bibinfo {title} {Magnetism in iron and nickel},\ }\href
  {https://doi.org/10.1103/PhysRevB.25.5766} {\bibfield  {journal} {\bibinfo
  {journal} {Phys. Rev. B}\ }\textbf {\bibinfo {volume} {25}},\ \bibinfo
  {pages} {5766} (\bibinfo {year} {1982})}\BibitemShut {NoStop}%
\bibitem [{\citenamefont {Bogolyubov}\ and\ \citenamefont
  {Tyablikov}(1959)}]{Bogolyubov59}%
  \BibitemOpen
  \bibfield  {author} {\bibinfo {author} {\bibfnamefont {N.~N.}\ \bibnamefont
  {Bogolyubov}}\ and\ \bibinfo {author} {\bibfnamefont {S.~V.}\ \bibnamefont
  {Tyablikov}},\ }\bibfield  {title} {\bibinfo {title} {Retarded and advanced
  {G}reen functions in statistical physics},\ }\href@noop {} {\bibfield
  {journal} {\bibinfo  {journal} {Sov. Phys. Dokl.}\ }\textbf {\bibinfo
  {volume} {4}},\ \bibinfo {pages} {589} (\bibinfo {year} {1959})}\BibitemShut
  {NoStop}%
\bibitem [{\citenamefont {K\"ormann}\ \emph {et~al.}(2009)\citenamefont
  {K\"ormann}, \citenamefont {Dick}, \citenamefont {Hickel},\ and\
  \citenamefont {Neugebauer}}]{Kormann09}%
  \BibitemOpen
  \bibfield  {author} {\bibinfo {author} {\bibfnamefont {F.}~\bibnamefont
  {K\"ormann}}, \bibinfo {author} {\bibfnamefont {A.}~\bibnamefont {Dick}},
  \bibinfo {author} {\bibfnamefont {T.}~\bibnamefont {Hickel}},\ and\ \bibinfo
  {author} {\bibfnamefont {J.}~\bibnamefont {Neugebauer}},\ }\bibfield  {title}
  {\bibinfo {title} {Pressure dependence of the {C}urie temperature in bcc iron
  studied by {\sl ab initio} simulations},\ }\href
  {https://doi.org/10.1103/PhysRevB.79.184406} {\bibfield  {journal} {\bibinfo
  {journal} {Phys. Rev. B}\ }\textbf {\bibinfo {volume} {79}},\ \bibinfo
  {pages} {184406} (\bibinfo {year} {2009})}\BibitemShut {NoStop}%
\end{thebibliography}%

\end{document}


\title{Supplemental Material for ``Origin of the suppression of magnetic order in MnSi under hydrostatic pressure''}

\author{P. Dalmas de R\'eotier}
\affiliation{Universit\'e Grenoble Alpes, CEA, Grenoble INP, IRIG-PHELIQS, F-38000 Grenoble, France}
\author{A. Yaouanc}
\affiliation{Universit\'e Grenoble Alpes, CEA, Grenoble INP, IRIG-PHELIQS, F-38000 Grenoble, France}
\author{D. Andreica}
\affiliation{Faculty of Physics, Babes-Bolyai University, 400084 Cluj-Napoca, Romania}
\author{R. Gupta}
\affiliation{Laboratory for Muon-Spin Spectroscopy, Paul Scherrer Institute,
CH-5232 Villigen-PSI, Switzerland}
\affiliation{Department of Physics, Indian Institute of Technology Ropar, Rupnagar, Punjab 140001, India}
\author{R. Khasanov}
\affiliation{Laboratory for Muon-Spin Spectroscopy, Paul Scherrer Institute,
CH-5232 Villigen-PSI, Switzerland}
\author{G. Lapertot}
\affiliation{Universit\'e Grenoble Alpes, CEA, Grenoble INP, IRIG-PHELIQS, F-38000 Grenoble, France}

\date{\today}

\begin{abstract}

Here, complementary information is presented, focusing on six aspects: (i) the crystal structure, (ii) the determination of $T_{\rm c}(p)$ from weak transverse-field $\mu$SR measurements, (iii) a discussion of zero-field measurements as a function of pressure or temperature and the fitting procedure, (iv) alternative models for the spectra recorded at 1.25~GPa, (v) the model for $m(T)$, and (vi) a description of the scaling law used to describe $J(p)$.
\end{abstract}

\maketitle

\section{Crystal structure}

MnSi crystallizes in the B20 structure with cubic space group $P2_13$ \cite{Boren33}. The Mn and Si atoms each occupy a $4a$ Wyckoff position located on a three-fold symmetry axis. The coordinates of the four sites are given in Table~\ref{Wyckoff}. For reference, the Mn (Si) position corresponds to $x$ = $x_{\rm Mn}$ = 0.138 ($x$ = $x_{\rm Si}$ = 0.845).

\begin{table}
  \caption{Coordinates of the four sites of Wyckoff position $4a$ in space group $P2_13$. The coordinates are expressed in units of the cubic lattice parameter.}\label{Wyckoff}
  \begin{tabular}{l@{\hskip 5mm}l} \hline\hline
site & coordinates \cr\hline
{\rm I} & $(x,x,x)$\cr
{\rm II} & $(\bar{x}+\frac{1}{2}, \bar{x}, x+\frac{1}{2})$\cr
{\rm III} & $(\bar{x}, x+\frac{1}{2}, \bar{x}+\frac{1}{2})$\cr
{\rm IV} & $(x+\frac{1}{2}, \bar{x}+\frac{1}{2}, \bar{x})$ \\[1mm]\hline\hline
  \end{tabular}
\end{table}

\section{Determination of $T_{\rm c}(p)$}
The magnetic transition temperature $T_{\rm c}$ was determined at each pressure from weak transverse-field (wTF) measurements \cite{Yaouanc11,Amato24}. This technique consists in applying a magnetic field ${\bf B}_{\rm ext}$ perpendicular to the initial polarization of the muons. Its magnitude $B_{\rm ext}$ is chosen such that it is lower than or on the order of the magnitude $B_{\rm loc}$ of the local field probed by muons implanted in MnSi in the absence of external field. For reference, $B_{\rm loc} \gtrsim 100$~mT at low temperature \cite{Takigawa80,Dalmas16}.

\begin{figure}
\includegraphics[width=\linewidth]{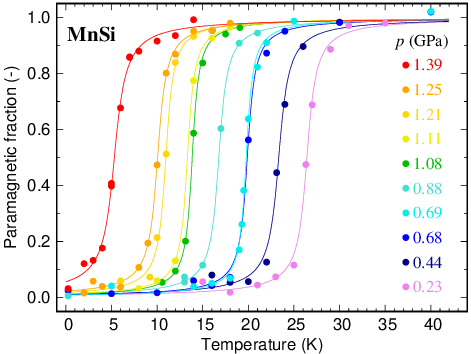}
\caption{Temperature dependence of the muon precessing fraction measured in wTF measurements at 5 mT for different hydrostatic pressures as indicated in the plot. The contribution from the pressure cell is subtracted. The precessing fraction corresponds to muons implanted in a non-magnetic environment. The full lines depict the results of fits with the phenomenological function $1/2+(1/\pi)\arctan[(T-T_{\rm c})/w)]$. }
\label{para_fraction}
\end{figure}

Let us consider in turn muons implanted in the paramagnetic phase of MnSi or in the walls of the pressure cell in the one hand and muons implanted in the magnetically ordered phase of MnSi, in the other hand. The former muons are submitted to a field approximately equal to $B_{\rm ext}$ and therefore precess at angular frequency $\approx\gamma_\mu B_{\rm ext}$. The latter muons are submitted to field ${\bf B}_{\rm ext} + {\bf B}_{\rm loc}$. Due both to the presence of magnetic domains and to the incommensurability of the magnetic order, ${\bf B}_{\rm loc}$ is not oriented in a specific direction relative to ${\bf B}_{\rm ext}$. As a consequence the muons are submitted to a distribution of fields, which makes them quickly depolarized. Therefore they do not exhibit coherent precession, in contrast to muons probing a paramagnetic phase. Figure~\ref{para_fraction} displays the results of the wTF measurements indicating sharp transitions over the whole pressure range and confirming that MnSi is magnetic within the whole sample volume at low temperature \cite{Andreica10}. Table~\ref{Tc} gathers the values of $T_{\rm c}$ at the different pressures. The results are consistent with those of the literature \cite{Pfleiderer97,Fak05}.
\begin{table}
  \caption{Magnetic transition temperature for the investigated pressures, resulting from the fit shown in Fig.~\ref{para_fraction}. Parameter $w$ characterizes the width of the magnetic transition. }\label{Tc}
  \begin{tabular}{l l l} \hline\hline
 $p$ (GPa) & $T_{\rm c}$ (K)   & $w$ (K)  \cr \hline
 0.23      & 26.42\,(7)  & 0.76\,(13) \cr
 0.44      & 23.33\,(13) & 0.82\,(20) \cr
 0.68      & 19.89\,(6)  & 0.63\,(15) \cr
 0.69      & 19.76\,(5)  & 0.54\,(7)  \cr
 0.88      & 16.77\,(4)  & 0.69\,(8)  \cr
 1.08      & 13.85\,(4)  & 0.55\,(8)  \cr
 1.11      & 13.32\,(4)  & 0.58\,(4)  \cr
 1.21      & 10.96\,(6)  & 0.63\,(11) \cr
 1.25      & 10.06\,(4)  & 0.72\,(6)  \cr
 1.39      & 5.32\,(7)   & 0.95\,(11) \cr \hline\hline
  \end{tabular}
\end{table}

\section{Zero-field $\mu$SR spectra}

\subsection{Pressure dependence at base temperature}

Zero-field (ZF) spectra recorded at different pressures are displayed in Fig.~2 together with the results of fits using the model described in the main text and common $P_Z^{\rm pc}(t)$ parameters for each type of pressure cells. In a first instance the muon localization site at Wyckoff position $4a$ of space group $P2_13$, i.e.\ $(x_\mu,x_\mu,x_\mu)$, was allowed to deviate from its low pressure value $x_\mu$ = 0.532 \cite{Amato14,Dalmas16,Dalmas18}. Since no definite trend of variation was noticed as a function of pressure, $x_\mu$ was fixed to 0.532 in the subsequent fits. On the same footing and for the same reason, parameter $r_\mu H/4\pi$ characterizing the muon Fermi contact interaction was fixed to its low pressure value $-1.04$ \cite{Amato14,Dalmas16,Dalmas18}.

For all the spectra shown in Fig.~2, the model assumption is ${\bf k} \parallel \langle 111\rangle$. Note the clear misfit around 0.06~$\mu$s for the spectra recorded at 1.21, 1.25, and 1.39~GPa. This point is further discussed in the main text and in Sec.~\ref{2comp}.

\subsection{Temperature dependence at various pressures}

\begin{figure}
\includegraphics[width=\linewidth]{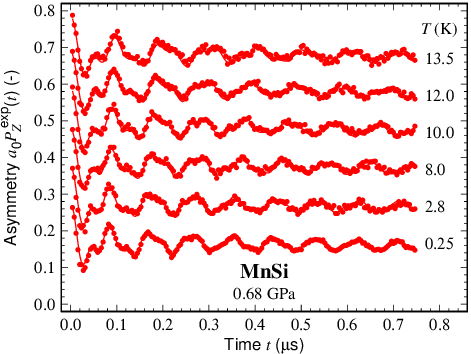}
\caption{ZF-$\mu$SR spectra recorded at 0.68~GPa for different temperatures as indicated in the plot. The full lines represent a combined fit to the whole data set with Eq.~6 and common parameters $x_\mu$, $r_\mu H /4\pi$, $\sigma_1$ and $\sigma_2$. The values for the last two parameters are given in Table~\ref{parameters}. The spectra recorded at temperatures of 2.8~K and above are vertically shifted for clarity.}
\label{ZF_6p8kbar_T}
\end{figure}

\begin{figure}
\includegraphics[width=\linewidth]{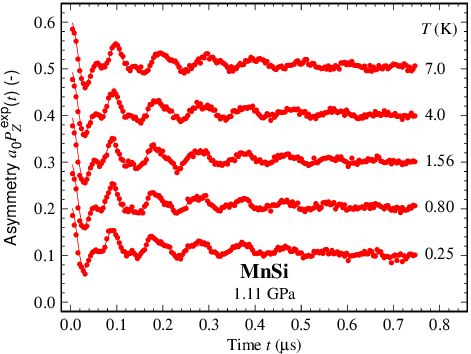}
\caption{Spectra similar to those of Fig.~\ref{ZF_6p8kbar_T} except that the pressure is 1.11~GPa.}
\label{ZF_11p1kbar_T}
\end{figure}

\begin{figure}
\includegraphics[width=\linewidth]{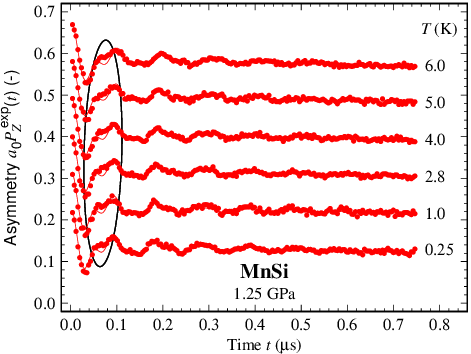}
\caption{Spectra similar to those of Fig.~\ref{ZF_6p8kbar_T} except that the pressure is 1.25~GPa.
  Note the systematic misfit around 0.06~$\mu$s.}
\label{ZF_12p5kbar_T}
\end{figure}

The Heisenberg exchange constant $J$ at each pressure is determined from the temperature dependence of the manganese magnetic moment. The sets of spectra from which $m(T)$ is derived are displayed in Fig.~\ref{ZF_6p8kbar_T}, \ref{ZF_11p1kbar_T} and \ref{ZF_12p5kbar_T} for $p$ = 0.68, 1.11, and 1.25~GPa, respectively. The full lines depict the results of fits using the model described in the main text. Similar to the case of the fits shown in Fig.~2, no remarkable dependence of parameters $x_\mu$ and $r_\mu H/4\pi$ was noticed in preliminary fits in which these parameters were permitted to freely vary. They were subsequently fixed to their low pressure value. In the same vein, no definite trend was observed for parameter $\sigma_2$, which, in the final fits, was forced to take a common value for each of the temperatures at a given pressure. The parameters resulting from the fits are reported in Table~\ref{parameters}.
\begin{table}
  \caption{Pressure dependence of MnSi physical parameters: $a_{\rm latt}$, $k$,\footnote{The $\mu$SR spectra are weakly sensitive to the magnitude of $a_{\rm latt}$ and $k$ \cite{Dalmas18}. Accordingly, the values used in the fits were fixed from neutron diffraction results \cite{Fak05,Grigoriev06}. Parameter $\sigma_1$ was computed using Eq.~2.} $\sigma_1$, $\sigma_2$, $T_{\rm he}$, and $J$. The twist and canting angles $\omega$ and $\Gamma$ of the ${\bf k} \parallel \langle 111 \rangle$ helimagnetic structure which directly derive from $\sigma_1$ and $\sigma_2$ are provided \cite{Chizhikov12,Dalmas24}. Parameter $\sigma_2$ at $p$ = 1.25~GPa is not displayed because it somewhat depends on the model used for the fits to the data. The values of $\omega$ are non-negligible compared to the regular rotation of magnetization between adjacent ferromagnetic planes \cite{Dalmas24}.}
  \label{parameters}
  \begin{tabular}{l  l  l  l  l} \hline\hline
$p$ (GPa) & 0  & 0.68\,(2) & 1.11\,(2) & 1.25\,(2)\cr
$a_{\rm latt}$ (nm) & 0.4558 & 0.4551 & 0.4546 & 0.4544 \cr
$k$ (nm$^{-1}$)    & 0.345 & 0.348 & 0.363 & 0.380 \cr
$\sigma_1$ (-) & 0.236 & 0.237 & 0.247 & 0.259 \cr
$\sigma_2$ (-) & 0.0179\,(3) &  0.011\,(4) & 0.026\,(2) & ---\cr
$\omega$ (deg.) & 0.87\,(1) & 0.6\,(2) & 1.1\,(1) & --- \cr
$\Gamma$ (deg.) & 0.42\,(1) & 0.3\,(1) & 0.6\,(1) & --- \cr
$T_{\rm he}$ (K) & 50.0\,(4) & 38.0\,(4) & 25.4\,(6) & 21.1\,(3) \cr
$J$ (meV) & 5.52\,(4) & 4.52\,(5) & 3.21\,(8) & 2.69\,(4) \cr\hline\hline
\end{tabular}
\end{table}

\begin{figure}
\includegraphics[width=\linewidth]{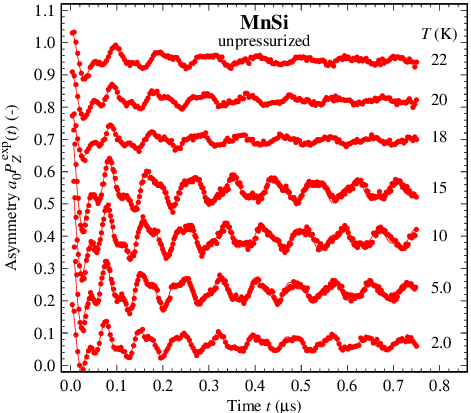}
\caption{Same caption as for Fig.~\ref{ZF_11p1kbar_T} except that the data are recorded with an unpressurized crystal and are those already published in Ref.~\onlinecite{Yaouanc20}.}
\label{ZF_0kbar_T}
\end{figure}

For completeness we have reanalyzed the low pressure data published in Ref.~\onlinecite{Yaouanc20} with the model described in the main text. The results are shown in Fig.~\ref{ZF_0kbar_T}. The value for $\sigma_2$ resulting from the combined fit is $\sigma_2$ = 0.0179\,(3) (Table \ref{parameters}). It compares favorably with the value $\sigma_2$ = 0.017\,(3) derived from the combined fit to the data recorded at 2~K in ZF and in a 0.3~T field applied along each of the principal directions of the cubic crystal structure \cite{Dalmas24}. We note that the accuracy on $\sigma_2$ is significantly improved. 

\section{Alternative models for the spectra recorded at 1.25~GP\lowercase{a}}
\label{alt_mod}

While the model used for the description of the ZF spectra presented in this report provides an excellent description of most of the available data (see, e.g.\ Figs.~2, \ref{ZF_6p8kbar_T}, and \ref{ZF_11p1kbar_T}), small but clear misfits are observed before 0.1~$\mu$s at 1.21~GPa and at higher pressures (Figs.~2 and \ref{ZF_12p5kbar_T}). It could be argued that the systematic misfit challenges the reliability of $m(T)$ as displayed in Fig.~4 for $p$ = 1.25~GPa. In this section we address this question by considering two alternative models and examine the resulting $T_{\rm he}$ values. It will be shown that both models support the conclusion presented in the main text.

Assuming the change in the spectral shape to be the signature of the progressive reorientation of the propagation vector from $\langle 111 \rangle$ to $\langle 001 \rangle$ evidenced in small angle neutron scattering experiments \cite{Bannenberg19}, we first consider a two-component model. In a second stage we deal with a phenomenological fit of the data.

\subsection{Two-component model} 
\label{2comp}

\begin{figure}
\includegraphics[width=\linewidth]{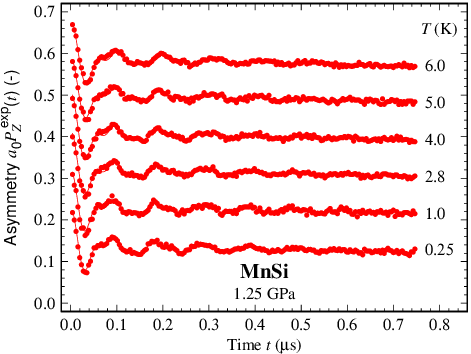}
\caption{Same spectra as those of Fig.~\ref{ZF_12p5kbar_T}. Here the lines represents a fit with, besides the $P_Z^{\rm pc}(t)$ function representing the pressure cell, a two-component model. The two components are respectively characterized by ${\bf k} \parallel \langle 111 \rangle$ and ${\bf k} \parallel \langle 001 \rangle$.  The former misfit observed around 0.06~$\mu$s has disappeared. More details about the fit function and the parameters are available in the text.
}
\label{ZF_13kbar_T_2comp}
\end{figure}

Here we substitute $P_Z^{\rm he}(t)$ of the main text with the sum of two components, each corresponding to helical orders with ${\bf k} \parallel \langle 111 \rangle$ and ${\bf k} \parallel \langle 001 \rangle$. The fit provides a good account of the data as displayed in Fig.~\ref{ZF_13kbar_T_2comp}. The fit parameters are as follows. The volume fraction of the former phase is approximately twice that of the latter. The magnitude $m$ of the manganese magnetic moment and the $\sigma_2 = 0.028\,(2)$ parameter are common to the two phases. The muon related parameters $x_\mu$ = 0.532 and $r_\mu H/4\pi$ = $-1.04$ keep their usual value for the ${\bf k} \parallel \langle 111 \rangle$ phase, while a small change of $x_\mu$ is found for the ${\bf k} \parallel \langle 001 \rangle$ phase [$x_\mu$ = 0.518\,(1), instead of $x_\mu$ = 0.532 for ${\bf k} \parallel \langle 111 \rangle$]. Such a small change may be the signature of an evolution in the crystal structure, which has not been reported so far, to the best of our knowledge. Alternatively it could reflect a small change in the electronic density. Finally and most importantly for our immediate interest, the temperature dependence $m(T)$ is very similar to that plotted in Fig.~4, with a consistent value $T_{\rm he}$ = 20.5\,(3)\,K.

\subsection{Phenomenological fits using the sum of two cosine functions}

We consider the model function
\begin{eqnarray}
  P_Z^{\rm he}(t) & = & \sum_{i=1}^2 p_i\left[\frac{2}{3} \cos(2\pi\nu_it + \varphi_i)\exp(-\lambda_{X,i}t) \right. \cr
               & & \hspace{10mm} \left. + \frac{1}{3}\exp(-\lambda_{Z,i}t)\right],
\label{phenomenological}
\end{eqnarray}
with $p_1+p_2$ = 1. This function was already used in a previous study of MnSi \cite{Kadono90}. It is purely phenomenological since the initial phases $\varphi_i$ are found to differ from 0, contrary to expectation for ZF spectra. The fact that the sum of two cosine functions with zero initial phases cannot provide a proper fit to the data can be understood from the shape of the Fourier transform of a $\mu$SR spectrum, where, among other features, a finite spectral weight is present between $\approx 100$ and 200~mT, in field units; see, e.g.\ Ref.~\onlinecite{Dalmas16}. We note that the decomposition of Eq.~\ref{phenomenological} in terms of 2/3 of precessing signal and 1/3 of relaxing signal, reminiscent of experiments performed on polycrystalline samples, also pertains for ZF measurements with a single crystal of MnSi due to the presence of K domains \cite{Dalmas16}.

\begin{figure}
\includegraphics[width=\linewidth]{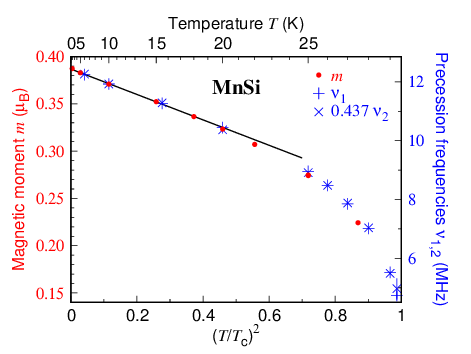}
\caption{Temperature dependence of the manganese magnetic moment at low pressure shown in Fig.~4 and the two muon spin spontaneous precession frequencies $\nu_1$ and $\nu_2$ published in Table I of Ref.~\onlinecite{Kadono90}. The full line depicts the result of the fit also shown in Fig.~4. Note that the fit with two precession frequencies is only phenomenological as the phases at time 0 do not vanish as they should do in a physically sound model. }
\label{ZF_m_nu_T2}
\end{figure}

In Fig.~\ref{ZF_m_nu_T2} we compare the temperature variation of $m$ at low pressure already shown in Fig.~4 of the main text with $\nu_1$ and $\nu_2$ as published in Ref.~\onlinecite{Kadono90}. The horizontal scale is proportional to  $T^2$ as in Fig.~4. The ratio $\nu_1/\nu_2$ is found independent of the temperature and $\nu_1(T)$ perfectly scales with $m(T)$ in the whole temperature range with a conversion factor equal to 32.15~MHz/$\mu_{\rm B}$. This suggests that $\nu_1(T)$ and $\nu_2(T)$ are good indicators of $m(T)$.

\begin{figure}[h]
\includegraphics[width=\linewidth]{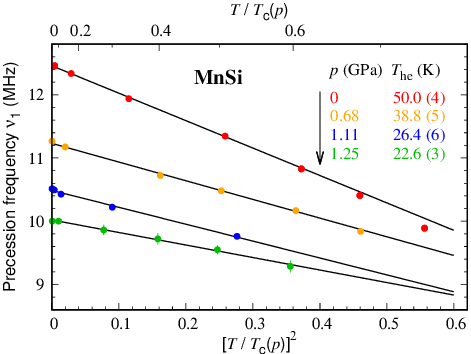}
\caption{Temperature variation of the lower precession frequency $\nu_1$ resulting from a fit to the data at the different pressures. The 0~GPa data are those of Ref.~\onlinecite{Yaouanc20}. The higher precession frequency $\nu_2$ is not shown for clarity. The full lines show the results of fits with function $\nu_1(T)=\nu_1(T=0)\left[1-(T/T_{\rm he})^2\right]$. The $T_{\rm he}$ values are given in the graph.}
\label{ZF_p_m_nu_T2}
\end{figure}

In view of Fig.~\ref{ZF_m_nu_T2}, it is tempting to reanalyze the $\mu$SR spectra recorded under pressure using Eq.~\ref{phenomenological}. We find that $\nu_2(T)$ still scales with $\nu_1(T)$ and we display $\nu_1(T)$ in Fig.~\ref{ZF_p_m_nu_T2}. Interestingly, the decay of $\nu_1$ is still proportional to the square of the temperature as $m(T)$ is (see Fig.~4), and parameters $T_{\rm he}$ controlling the decays of $\nu_1(T)$ and $m(T)$ at the different pressures are very similar. This is an additional indication that $T_{\rm he}$ at 1.25~GPa is reliably determined, despite the presence of magnetic phases with different ${\bf k}$ vectors.

\section{Quadratic temperature decay of the magnetic moment $m$}
\label{quadratic}
We evaluate the reduction of $m$ arising from the thermal excitation of spin waves. In a helimagnet like MnSi, the spin wave are quantized in so-called helimagnons. A gapless mode is derived for these helimagnons \cite{Maleyev06,Belitz06}. In linear spin-wave theory, its dispersion relation is
\begin{eqnarray}
\omega_{\bf q}^2 & = & c_\parallel q_\parallel^2 + c_\perp q_\perp^4,
\label{dispersion}
\end{eqnarray}
where ${\bf q}_{\parallel,\perp}$ are the wavevector components respectively parallel and perpendicular to ${\bf k}$ and $c_\parallel$ and $c_\perp$ are elastic constants that satisfy the relations
\begin{eqnarray}
c_\parallel  =  \left (\frac{B_1k S}{ \hbar} \right )^2  {\rm and} \,\,\,  c_\perp =  \epsilon \frac{c_\parallel}{k^2},
\label{elastic}
\end{eqnarray}
where $\epsilon$ = 1/2 in the simplest approximation. Here $B_1$ is the stiffness constant appearing in the continuum expression of the magnetic energy
\begin{eqnarray}
{\mathcal E} 
&  = & V \int \left [\frac{B_1 q^2}{2} |{\bf S}_{\bf q}|^2  
  + i D {\bf q} \cdot ({{\bf S}}_{\bf q} \times {{\bf S}}_{-{\bf q}})  \right]
\frac{{\rm d}^3{\bf q}}{(2\pi)^3}, \cr & &
\label{class_dynamics_9}
\end{eqnarray}
which is valid at small wavevector \cite{Bak80}. Parameter $V$ denotes the sample volume.

Equation \ref{dispersion} has the expected cylindrical symmetry for a lamellar structure \cite{Chaikin95} with an energy depending quadratically on the wavevector for spin waves propagating perpendicular to ${\bf k}$, i.e.\ in ferromagnetic planes, and a linear dependence for modes parallel to ${\bf k}$, a direction along which the magnetic structure is antiferromagnetic. Note that Eq.~\ref{dispersion} is approximate as its right-hand side must necessarily involve an additional $q_\parallel^2$ term so as to stabilize the magnetic order and avoid a Landau--Peierls instability \cite{Garst17}. The presence of such a term, even small, ensures the mathematical convergence of the integral introduced thereafter (Eq.~\ref{int_BZ}) \cite{Dalmas21}.

Inspired by the $T^{3/2}$ Bloch's law for the deviation of the spontaneous moment of a ferromagnet from its $T=0$ value  which is assigned to the excitations of magnons \cite{*[{see, e.g., }] [{.}] vanKranendonk58},
\begin{eqnarray}
m(T) & = & m(T=0)\left(1 -\frac{1}{N}\sum_{\bf q} n_{\bf q}\right).
\end{eqnarray}
Here $N$ is the number of magnetic ions in the crystal, ${\bf q}$ is a vector of the first Brillouin zone and $n_{\bf q}$ is the population of helimagnons with wavevector ${\bf q}$. Switching for convenience to the continuum limit,
\begin{eqnarray}
\sum_{\bf q} n_{\bf q} & = & V \int \left[\exp(\hbar\omega_{\bf q}/k_{\rm B}T)-1\right]^{-1}\,\frac{{\rm d}^3{\bf q}}{(2\pi)^3}.
\label{int_BZ}
\end{eqnarray}
The evaluation of the integral leads to Eq.~4 of the main text with
\begin{eqnarray}
T_{\rm he} = 4\sqrt{\frac{3}{\pi v_0 }}\frac{\hbar}{k_{\rm B}}  \left (c_\parallel c_\perp \right)^{1/4},
\label{moment_4}
\end{eqnarray}
where $v_0$ = $a_{\rm latt}^3/4$ is the volume per magnetic ion. Equation~5 is found to derive from the preceding equality, Eq.~\ref{elastic}, and the relation \cite{Chizhikov12}
\begin{eqnarray}
J & = & \frac{2 B_1}{3a^2_{\rm latt}}.
\label{Chizhikov12}
\end{eqnarray} 

\section{Scaling law for $J(p)$}
\label{scaling}

The following relations between $T_{\rm c}$ and the energy $\hbar \omega_{\bf q}$ of a spin wave at wavevector ${\bf q}$ 
\begin{eqnarray}
k_{\rm B} T_{\rm c} (p) \propto m^2(p) \sum_{\bf q} \omega_{\bf q}(p),
\label{scaling_1}
\end{eqnarray}
and
\begin{eqnarray}
k_{\rm B} T_{\rm c} (p) \propto m^2(p) \left[ \sum_{\bf q} \omega_{\bf q}^{-1}(p)\right]^{-1},
\label{scaling_2}
\end{eqnarray}
have been derived, respectively in the mean-field \cite{*[{See, e.g., }] [{}] Wang82} and in the random-phase approximation \cite{Bogolyubov59}. As before, the sums are over the Brillouin zone vectors. Although exact relations are available in the literature we have introduced the proportionality symbol in Eqs.~\ref{scaling_1} and \ref{scaling_2}. As explained in Ref.~\onlinecite{Kormann09}, while the absolute value of $T_{\rm c}$ is rather sensitive to the approximations used to solve the Heisenberg model Hamiltonian, its pressure dependence is well approximated by these equations.

Now, in general, the spin-wave energy is proportional to the exchange constant. This can be directly checked for the helimagnon dispersion relation (Eq.~\ref{dispersion} with Eqs.~\ref{elastic} and \ref{Chizhikov12}). 
With this information at hand, we obtain the scaling relation
\begin{eqnarray}
k_{\rm B} T_{\rm c} (p) \propto  J(p) m^2(p),
\label{scaling_3}
\end{eqnarray}
introduced in the main text. 

\bibliography{reference}